\documentclass[twocolumn,aps,showpacs,prb,eqsecnum]{revtex4-1}
\usepackage[latin9]{inputenc}
\usepackage{amsmath}
\usepackage{amssymb}
\usepackage{graphicx}
\usepackage{esint}
\usepackage{epsf}

\makeatletter
\@ifundefined{textcolor}{}
{%
 \definecolor{BLACK}{gray}{0}
 \definecolor{WHITE}{gray}{1}
 \definecolor{RED}{rgb}{1,0,0}
 \definecolor{GREEN}{rgb}{0,1,0}
 \definecolor{BLUE}{rgb}{0,0,1}
 \definecolor{CYAN}{cmyk}{1,0,0,0}
 \definecolor{MAGENTA}{cmyk}{0,1,0,0}
 \definecolor{YELLOW}{cmyk}{0,0,1,0}
}

\usepackage{epsf}\usepackage{dcolumn}\usepackage{bm}

\newcommand{\bk}{{\bf k}}
\newcommand{\bq}{{\bf q}}
\newcommand{\bp}{{\bf p}}
\newcommand{\bkp}{{\bf k}'}
\newcommand{\bpp}{{\bf p}'}
\newcommand{\bv}{{\bf v}}
\newcommand{\ek}{\varepsilon_{\mathbf k}}
\newcommand{\ekp}{\varepsilon_{{\mathbf k}'}}
\newcommand{\ep}{\varepsilon_{\mathbf p}}

\newcommand{\ti}{\tau_{\mathrm{i}}}
\newcommand{\tis}{\tau_{\mathrm{i}1} }
\newcommand{\tid}{\tau_{\mathrm{i}2} }

\newcommand{\beq}{\begin{equation}}
\newcommand{\eeq}{\end{equation}}
\newcommand{\bea}{\begin{eqnarray}}
\newcommand{\eea}{\end{eqnarray}}
\newcommand{\otau}{1/\tau_{\mathrm{ee}}}
\def\blfootnote{\xdef\@thefnmark{}\@footnotetext}

\makeatother

\begin{document}
\title{Resistivity of non-Galilean-invariant Fermi- and non-Fermi liquids}

\author{H. K. Pal$^{1}$, V. I. Yudson$^{2}$, and D. L. Maslov$^{1}$ 
}
\affiliation{ $^{1}$Department of Physics, University of Florida, Gainesville,
FL 32611-8440, USA\\
 $^{2}$Institute for Spectroscopy, Russian Academy of Sciences, Troitsk,
Moscow Region, 142190, Russia}
\date{\today}
\begin{abstract}

While it is well-known that the electron-electron (\emph{ee}) interaction
cannot affect the resistivity of a Galilean-invariant Fermi liquid
(FL), the reverse statement is not necessarily true: the resistivity
of a non-Galilean-invariant FL does not necessarily follow a $T^{2}$
behavior. The $T^{2}$ behavior is guaranteed only if Umklapp processes
are allowed; however, if the Fermi surface (FS) is small or the electron-electron
interaction is of a very long range, Umklapps are suppressed.
In this case, a $T^{2}$ term can result only from a combined--but
distinct from 
 quantum-interference corrections-- effect of the electron-impurity
and \emph{ee} interactions. Whether the $T^{2}$ term is present depends
on 1) dimensionality {[}two dimensions (2D) vs three dimensions
(3D){]}, 2) topology (simply- vs multiply-connected), and 3) shape (convex vs concave) of the FS. In particular, the $T^{2}$ term is absent
for any quadratic (but not necessarily isotropic) spectrum both in
2D and 3D. The $T^{2}$ term is also absent for a convex and simply-connected
but otherwise arbitrarily anisotropic FS in 2D. The origin of this
nullification is approximate integrability of the electron motion on a
2D FS, where the energy and momentum conservation laws do not allow
for current relaxation to leading --second--order in $T/E_{F}$ ($E_{F}$
is the Fermi energy). If the $T^{2}$ term is nullified by the conservation
law, the first non-zero term behaves as $T^{4}$. The same applies to a quantum-critical metal in the vicinity of a Pomeranchuk instability, with a proviso that the leading (first non-zero) term in the resistivity scales as $T^{\frac{D+2}{3}}$
($T^{\frac{D+8}{3}}$).
We discuss a number
of situations when integrability is weakly broken, e.g., by inter-plane hopping
in a quasi-2D metal or by warping of the FS as in the surface states of
topological insulators of the Bi$_{2}$Te$_{3}$ family.
The paper is intended to be self-contained and pedagogical; 
review of the existing results is included along with the original
ones wherever deemed necessary for completeness.
\end{abstract}

\maketitle
\section{\label{sec1} Introduction}
\blfootnote{Submitted to a special issue of the Lithuanian Journal of Physics dedicated to the memory of Y. B. Levinson.}
A $T^{2}$ scaling of the resistivity with temperature ($T$) is considered as
an archetypal signature of the Fermi-liquid (FL) behavior in metals. This result
owes its origin to the Pauli exclusion principle which dictates that, at low temperatures,
only those quasiparticles that reside within a width of order $T$ near the Fermi energy
participate in binary collisions.
This argument, however, applies only to the inverse of the quasiparticle
relaxation time $1/\tau_{\mathrm{ee}}$ but not to the resistivity, $\rho$, {\em per
se}: the $T^2$ scaling of the former does not necessarily imply that of the latter. A very simple
example is a Galilean-invariant FL, where the electron-electron (\emph{ee}) interaction does not affect
the resistivity, although $1/\tau_{\mathrm{ee}}$, as measured, e.g., by thermal conductivity, does scale as $T^{2}$.
The reason is that, since velocities of electrons
are proportional to their respective momenta, conservation of momentum
automatically implies conservation of the electric current.
In order to achieve a steady-state current under the effect of an external
electric field, a momentum relaxation mechanism is needed. 

Of course, the FL of electrons in a metal is not Galilean-invariant.
In the presence of lattice, the current may be relaxed by Umklapp collisions,~\cite{landau:36} which conserve
the quasimomentum up to a reciprocal lattice vector: $\mathbf{k}+\mathbf{p}=\mathbf{k'}+\mathbf{p'}+\mathbf{b}$.
Umklapp processes are allowed, however, if the incoming electron
momenta $\mathbf{k}$ and $\mathbf{p}$ as well as the momentum transfer
$\mathbf{q}=\bk-\bkp=\bpp-\bp$ are all of order $\mathbf{b}$.
These requirements are satisfied
1) if the Fermi surface is large enough, e.g., at least quarter-filled
in the tight-binding case,~\cite{abrikosov} and 2) if the interaction is sufficiently short-ranged.
In conventional metals, these two conditions are easily met due
to a large number of carriers and effective screening of the Coulomb interaction; thus
Umklapp collisions occur at a rate comparable to $1/\tau_{\mathrm{ee}}$,
and $\rho\propto T^{2}$. 

However, there are situations when these
conditions are not met; e.g.,
the first condition is violated in systems with low carrier concentration,
such as degenerate semiconductors, semimetals, surface states
of three dimensional topological insulators, etc., and the second
condition is violated when a metal is tuned to the vicinity of a Pomeranchuk-type
quantum phase transition~\cite{pomeranchuk:58} (QPT), e.g,  a ferromagnetic QPT. A Pomeranchuk-type QPT is a $q=0$ instability of the ground state, manifested by a
divergence of long-wavelength fluctuations of the order parameter. The effective radius
of the interaction mediated by the exchange of such fluctuations diverges at the QPT.
One of the consequences of this divergence is the FL breakdown,
as manifested by a non-Fermi--liquid (NFL) scaling $1/\tau_{\mathrm{ee}}\propto T^{\gamma}$
with $\gamma\leq 1$,
but another one is the concurrent suppression of Umklapp processes. 

If Umklapps are suppressed (and the temperature is too low for the electron-phonon interaction to be effective), 
current can be relaxed only via electron-impurity ({\em ei}) collisions.
Still, the normal, i.e., momentum-conserving, {\em ee} collisions
can affect the resistivity, if certain conditions are met. The main purpose of this paper is to
summarize and analyze these conditions.
The combined effect of  normal
{\em ee} and \emph{ei} interactions does not necessarily lead to 
the $T^{2}$ dependence (or its NFL analog) of the resistivity. 
Whether this happens
depends on three factors: 1) dimensionality [two dimensions
(2D) vs three dimensions (3D)], 2) topology (simply vs multiply connected),
and 3) shape (convex vs concave) of the Fermi surface (FS). The
$T^{2}$ term is absent not only for a Galilean-invariant but, more generally, for an isotropic FL with a non-parabolic spectrum, as well as for anisotropic but quadratic spectrum.  In 2D,
the conditions are more stringent. In addition to cases mentioned above,
the $T^{2}$ term is absent for a simply-connected and convex but otherwise arbitrarily anisotropic FS.
The reason behind this is that the $T^{2}$ term
arises from electrons confined to move along the FS contour such that, for the convex case,
momentum and energy conservations are similar
to the 1D case, where no relaxation is possible.

The issue of an interplay between normal {\em ee} and {\em ei} interactions has a long history, and it is beyond the scope of this paper to give a comprehensive review of the existing literature; some aspects relevant to 3D metals are reviewed in Ref.~\onlinecite{bass:90}.  Very briefly, the first notion that normal processes can affect the resistivity even in a single-band metal probably goes back to the paper by Debye and Conwell. \cite{debye54}
There is also a large body of work on normal collisions in multi-band metals, following the original paper by Baber,~\cite{baber} both at the phenomenological (reviewed thoroughly in Ref.~\onlinecite{levinson}) and microscopic~\cite{appel78} levels.
That momentum relaxation occurs differently in 2D as compared to 3D was pointed by Gurzhi, Kopeliovich, and Rutkevich, first for the electron-phonon\cite{gurzhi_eph} and then for the {\em ee}\cite{gurzhi_ee} interactions. 
Maebashi and Fukuyama \cite{maebashi97_98} analyzed an interplay between normal and Umklapp collisions for an anisotropic 2D FS and found that the normal collisions do not give rise to a $T^2$ term as long as the FS is convex. Rosch and Howell (Ref.~\onlinecite{rosch05}a) and Rosch (Ref.~\onlinecite{rosch05}b) showed that a similar nullification happens for the $\omega^2$ term in the optical conductivity in a disorder-free 2D system. 
Chubukov and two of us (D. L. M. and V. I. Y.) generalized the analysis for a NFL near the Pomeranchuk QPT.~\cite{maslov1}  Scaling of the resistivity near a convex-to-concave transition was studied in Ref.~\onlinecite{pal}. 
This paper
expands on our recent works \cite{maslov1,pal} and provides some
more details.

It is worth noting that the effects studied in this paper occur already within the semiclassical theory of transport that neglects quantum
interference between {\em ee} and {\em ei} scatterings. Whether semiclassical description makes sense is one of the issues analyzed in the paper (cf. Sec.~\ref{sec:qint}): as a general rule, semiclassical effects can be considered separately from quantum-interference ones in the ballistic but not in the diffusive limit. Another effect not captured by the semiclassical Boltzmann equation (BE) is the viscous correction to the resistivity, discussed in Sec.~\ref{sec:visc}. The viscous correction
is expected to be the leading contribution to the resistivity resulting from the {\em ee} interaction in the high-temperature, hydrodynamic regime, when the {\em ee} mean free path is sufficiently short. 

The rest of the paper is organized as follows. We begin by formulating
the problem in terms of the Boltzmann equation (BE) both for the FL and the NFL cases
in Sec.~\ref{sec2}.
In Sec.~\ref{sec3},
we solve the BE  
perturbatively with respect to {\em ee} scattering, which is an adequate approximation at low enough temperatures,
and analyze various stituations mentioned above. 
In Sec.~\ref{sec4}, we discuss the opposite limit of high temperatures, when the {\em ee} contribution to the resistivity saturates, and show that a true scaling regime, with an appreciable difference between the low and high temperature limits of the resistivity, does not exist in a single-band metal (Sec.~\ref{sec:sat}). Such a regime is shown to exist for a two-band metal with very different masses (Sec.~\ref{sec:2band}).
In Sec.~\ref{sec5}, we analyze the limits of the validity of the results based on the semi-classical BE with respect to both quantum (Sec.~\ref{sec:qint})  and classical (Sec.~\ref{sec:visc}) correlations between {\em ee} and {\em ei} interactions.
Our concluding remarks are presented in
Sec.~\ref{sec6}.

\section{\label{sec2} Boltzmann equation: Generalities}

\subsection{Collision integral}

The most straightforward way to find the effect of the \emph{ee} interaction
on the conductivity in the semi-classical regime is via the Boltzmann
equation (BE) which, for the case of a time-independent and spatially-uniform
external electric field ${\bf E}$, reads 
\begin{equation}
-e{\mathbf{E}}\cdot\frac{\partial f_{\mathbf{k}}}{\partial\bk}=-I_{\mathrm{ei}}\left[f_{\mathbf{k}}\right]-I_{\mathrm{ee}}\left[f_{\mathbf{k}}\right],
\label{boltz}
\end{equation}
where 
$-e$ is the electron charge
and $f_{\mathbf{k}}$ is the distribution function.
The collision integrals $I_{\mathrm{ee}}$ and $I_{\mathrm{ei}}$
on the right-hand side describe the effects of the \emph{ee} and \emph{ei}
interactions, respectively. Explicitly,
\begin{equation}
I_{\mathrm{ei}}=
\int_{\bk'}
w_{\mathbf{k}^{\prime}\mathbf{k}}\left(f_{\mathbf{k}}-f_{\mathbf{k}^{\prime}}\right)\delta\left(\ek-\ekp\right),\label{iei}
\end{equation}
and
\begin{eqnarray}
I_{\mathrm{ee}} & = & \int_{\bp}
\int_{\bp'}
\int_{\bk'}
W_{\mathbf{k,p\rightarrow k}^{\prime}\mathbf{p}^{\prime}}\notag\\
& &\times\delta\left(\varepsilon_{\mathbf{k}}+\varepsilon_{\mathbf{p}}-\varepsilon_{\mathbf{k}^{\prime}}-\varepsilon_{\mathbf{p}^{\prime}}\right)\delta\left(\mathbf{k+p-k}^{\prime}-\mathbf{p}^{\prime}\right)\notag\\
 &  & \times\left[f_{\mathbf{k}}f_{\mathbf{p}}\left(1-f_{\mathbf{k}^{\prime}}\right)\left(1-f_{\mathbf{p}^{\prime}}\right)-f_{\mathbf{k}^{\prime}}f_{\mathbf{p}^{\prime}}\left(1-f_{\mathbf{k}}\right)\left(1-f_{\mathbf{p}}\right)\right],\notag\\
\label{iee1}
\end{eqnarray}
where $\int_{\bk}$ is a short-hand notation for $\int\frac{d\mathbf{k}}{(2\pi)^{D}}$, and $w_{\bk,\bkp}$ and $W_{\mathbf{k,p\rightarrow k}^{\prime}\mathbf{p}^{\prime}}$ are the \emph{ei} and \emph{ee} scattering probabilities, correspondingly. 
For a weak electric field, the left-hand
side of the BE reduces to $e\bv_{\bk}\cdot{\bf E}n_{\bk}'$, where
$\bv_{\bk}$ is the electron group velocity and $n_{\bk}\equiv n(\ek)$ is the
equilibrium distribution function, with prime denoting a derivative
with respect to the electron energy, $\ek$ (measured from the Fermi energy). Linearizing the {\em ee} collision integral on the right-hand side
with respect to the non-equilibrium correction to $n_{\bk}$, defined
as
\beq
f_{\mathbf{k}}=n_{\mathbf{k}}-Tn_{\mathbf{k}}^{\prime}g_{\mathbf{k}}=n_{\mathbf{k}}+n_{\mathbf{k}}\left(1-n_{\mathbf{k}}\right)g_{\mathbf{k}},\label{dist1}
\eeq
one obtains\cite{abrikosov}
\begin{eqnarray}
I_{\mathrm{ee}} & = &
 \int_{\bp}
 \int_{\bp'}
\int_{\bk'}
 W_{\mathbf{k,p\rightarrow k}^{\prime}\mathbf{p}^{\prime}}\nonumber \\
 &  & \times\left(g_{\mathbf{k}}+g_{\mathbf{p}}-g_{\mathbf{k}^{\prime}}-g_{\mathbf{p}^{\prime}}\right)n_{\mathbf{k}}n_{\mathbf{p}}\left(1-n_{\mathbf{k}^{\prime}}\right)\left(1-n_{\mathbf{p}^{\prime}}\right)\nonumber \\
 &  & \times\delta\left(\mathbf{k+p-k}^{\prime}-\mathbf{p}^{\prime}\right)\delta\left(\varepsilon_{\mathbf{k}}+\varepsilon_{\mathbf{p}}-\varepsilon_{\mathbf{k}^{\prime}}-\varepsilon_{\mathbf{p}^{\prime}}\right).\label{iee2}
\end{eqnarray}

\subsection{Pomeranchuk quantum criticality}
\label{sec:pom}
In addition to the case of a generic FL, we will be also interested in a special but widely studied case
of a FL near a Pomeranchuk-type QPT, \cite{pomeranchuk:58} which breaks the rotational symmetry and/or topology of the FS but leaves the translational symmetries intact. Examples of such a QPT include ferromagnetic and electronic nematic transitions. \cite{fradkin10} As opposed to, e.g., charge-density waves and antiferromagnets, both the ordered and disordered phases are spatially uniform, and the transition is manifested via the divergence of certain susceptibility at $q=0$. Therefore, critical fluctuations near the QPT are long-ranged, and the effective interaction among electrons, mediated by these fluctuations, is of a long range as well.  Since a FL is, in general, unstable with respect to long-range interaction, the quantum-critical region of the phase diagram near the QPT is characterized by manifestly non-Fermi liquid (NFL) properties, such as a divergence of the specific heat coefficient. However, the long-range nature of the effective interaction has another aspect; namely, small-angle scattering at critical fluctuations effectively prohibits Umklapp processes which, in the absence of disorder, are necessary to render the resistivity finite.~\cite{maslov1} Therefore, the {\em ee} contribution to the resistivity can result only from an interplay between {\em ei} and normal {\em ee} collisions.

In this Section, we briefly summarize the properties of the simplest model describing a QPT of the Pomeranchuk type: the Hertz-Moriya-Millis (HMM) model.~\cite{hmm} In this model, electrons are assumed to interact via an effective potential
proportional to the divergent susceptibility of the order parameter. Details of the effective interaction depend on whether instability occurs in the charge or spin channel but, for our purposes, it suffices to model the interaction by a scalar function
\beq
U_{\mathrm{eff}}\left(q,\omega\right)=\frac{\nu_F^{-1}}{\delta+a^{2}q^{2}-\frac{i\omega}{v_{F}q}},
\label{hm1}
\eeq
where $\nu_F$ is the density of states, $\delta>0$ is the \lq\lq distance\rq\rq\/  to the critical point along the axis of the control parameter (pressure, doping, etc.), and $a$ is the radius of interaction in the critical channel.  Since (\ref{hm1}) can be derived, strictly speaking, only in the random-phase approximation, one needs to require that $k_Fa\gg 1$ (Ref.~\onlinecite{dzero04}). [Alternatively, one can assume that the coupling between electrons and critical fluctuations is weak; \cite{pepin06} results of these two approaches differ only by re-definition of parameters.] The imaginary part of $U_{\mathrm{eff}}$ results from Landau damping of critical fluctuations by itinerant electrons. The correlation length of critical fluctuations $\xi=a/\sqrt{\delta}$ diverges at the QPT, where $\delta=0$.

 For the interaction in Eq.~(\ref{hm1}), the inverse quasiparticle lifetime (the imaginary part of the self-energy) behaves as $\otau\propto T^2$ for $T\ll T_{\mathrm{FL}}\equiv v_Fa^2/\xi^3$ and as $\otau\propto T^{D/3}$ for $T\gg T_{\mathrm{FL}}$. 
 The energy scale $T_{\mathrm{FL}}$ separates the FL and NFL regions of the phase diagram.
 Since the momentum transfers are small in both regions (${\bar q}\sim \xi^{-1}$ for $T\ll T_{\mathrm{FL}}$ and ${\bar q}\sim (\omega/v_Fa^2)^{1/3}\sim (T/v_Fa^2)^{1/3}$ for $T\ll T_{\mathrm{FL}}$), the transport scattering time is longer than the lifetime:  $1/\tau_{\rm ee}^{\rm tr}\sim \left(\otau\right)\left({\bar q}/k_F\right)^2$. In the NFL region, $1/\tau_{\rm ee}^{\rm tr}\propto T^{(D+2)/3}$; in the FL regime, $1/\tau_{\rm ee}^{\rm tr}\propto T^2$ with a small prefactor.~\cite{schofield:99} The conventional wisdom was  that \lq\lq transportization\rq\rq\/ of the relaxation time was the only manifestation of the long-range nature of the interaction, so that
 the resistivity could simply be obtained by substituting the transport time into the Drude formula.~\cite{schofield:99,anna:07} This yields $\rho\propto T^{5/3}$ and $\rho\propto T^{4/3}$  in the NFL regions in 3D and 2D, correspondingly. The $5/3$ scaling of the resistivity is indeed close to what has been observed experimentally
in a number of itinerant ferromagnets near a QPT.~\cite{exp_fm}  The reasoning tacitly assumes, however, that Umklapp collisions
are still present, and occur at a rate comparable to that for normal ones, so that the transport time for normal collisions 
gives a reasonable estimate for the Umklapp scattering time. As we have already pointed out at the beginning of this Section, this assumption is not satisfied for a long-range interaction. In the next Section, we will quantify this statement. Before we proceed further, some general comments on the HMM model are in order. 

First, we are going to use the BE even in the NFL region of the QPT, where quasiparticles are not well defined, i.e., when $\otau\gg T$. This seems to be inconsistent with the general criterion of the validity of the BE.~\cite{physkin} However, well-defined quasiparticles are not required for the BE to be valid in a special case, when the effective {\em ee} interaction  can be treated in the Migdal-Eliashberg approximation, i.e., when the self-energy depends on the electron energy but not the momentum and vertex corrections are small. In this case, the BE can be derived in the Keldysh technique without any conditions on the parameter $\tau_{\mathrm{ee}}T$, as long as a much weaker condition $\tau_{\mathrm{ee}}\varepsilon_F\gg 1$ is satisfied.~\cite{rammer}
This argument, formulated first by Prange and Kadanoff for the electron-phonon interaction,\cite{prange:64} was used later by a number of researchers in a wider context, \cite{graf:93} and also applicable to NFL systems, given that they allow Migdal-Eliashberg description. Having said that, we come to the second point, which is that the Migdal-Eliashberg treatment of the HMM model, thought previously to be controllable at least in the $1/N$ approximation,\cite{pepin06} has recently been shown to break down beyond the second loop order.~\cite{eliashberg_breakdown} While acknowledging this problem, we remark that the processes responsible for this breakdown are effectively 1D-like scattering events, in which both the initial and final fermions move along the same line. Although these processes are dangerous for the single-particle self-energy, their contribution to the conductivity should be reduced by at least the \lq\lq transport factor\rq\rq\/, which discriminates against small-angle scattering. 

Once the BE is adopted, the difference between the FL and NFL regimes becomes formal: the dependence of the {\em ee} scattering probability on the energy transfer may be neglected in the former but not in the latter. 

\subsection{Matrix elements on a lattice: normal vs Umklapp processes}

\begin{figure}[t]
\includegraphics[width=0.5\textwidth]{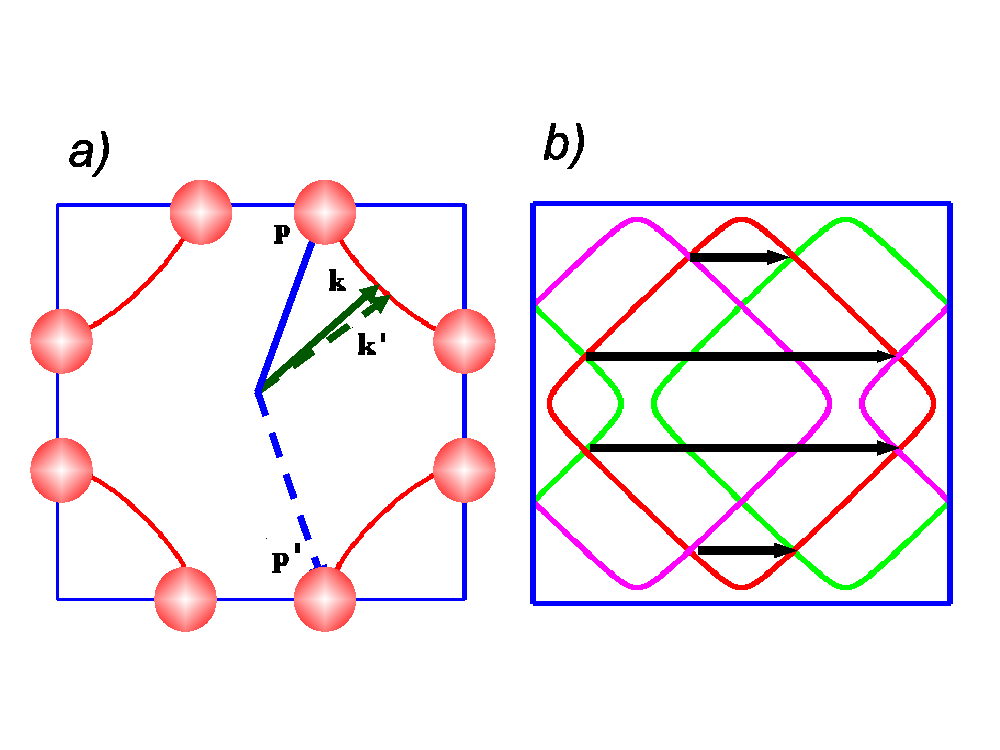}
\caption{({\em a}) Umklapp process for a long-range electron-electron interaction.  One of the electrons (with initial momentum $\bk$) is scattered by a small angle via the {\em ee} interaction,
while another one (with initial momentum $\bp$) is scattered by the lattice all the way across the Brillouin zone. {\em b})  
Umklapp processes for large momentum transfer. The original FS is in the center. Reprinted from Ref.~\onlinecite{maslov1}, courtesy of the APS.}
\label{fig:fig1}
\end{figure}
 The  interaction potential between electrons on a lattice 
$U\left(\mathbf{r}_{1},\mathbf{r}_{2}\right)$ depends on the coordinates
of two electrons separately rather than on their relative coordinate;
transforming to the center-of-mass
and relative coordinates,
$U\left(\mathbf{r}_{1},\mathbf{r}_{2}\right)=U\left(\mathbf{r}_{1}-\mathbf{r}_{2},\frac{\mathbf{r}_{1}+\mathbf{r}_{2}}{2}\right)$, where
$U$ is a periodic function
of $\frac{\mathbf{r}_{1}+\mathbf{r}_{2}}{2}$  but not
of $\mathbf{r}_{1}-\mathbf{r}_{2}.$ (The time-dependence of the effective interaction is not essential for the analysis below, and will be omitted.) Consequently, 
\beq
U\left(\mathbf{r}_{1},\mathbf{r}_{2}\right)=\int_\bq\sum_{\mathbf{b}}e^{-i\mathbf{q\cdot}\left(\mathbf{r}_{1}-\mathbf{r}_{2}\right)}e^{-i\mathbf{b\cdot}\left(\mathbf{r}_{1}+\mathbf{r}_{2}\right)/2}U\left(\mathbf{q,b}\right),
\eeq
where ${\bf b}$ is the reciprocal lattice vector and the volume of the system is put to unity. The matrix element of $U$ on the Bloch wave functions $\Psi_\bk\left(\mathbf{r}\right)=\sum_{\mathbf{b}}u_{\mathbf{k}}\left(\mathbf{b}\right)e^{i(\mathbf{k-b)\cdot r}}$ reads 
\begin{widetext} 
\begin{eqnarray}
M_{\mathbf{kp\rightarrow k}^{\prime}\mathbf{p}^{\prime}}
= \sum_{{\bf b},\mathbf{b}_{1}\dots
\mathbf{b}_{4}}\delta_{\mathbf{k-k}^{\prime}\mathbf{+b}_{1}-\mathbf{b}_{2}\mathbf{,p}^{\prime}-\mathbf{p+b}_{3}-\mathbf{b}_{4}}u_{\mathbf{k}^{\prime}}^{\ast}\left(\mathbf{b}_{1}\right)u_{\mathbf{k}}\left(\mathbf{b}_{2}\right)u_{\mathbf{p}^{\prime}}^{\ast}\left(\mathbf{b}_{3}\right)u_{\mathbf{p}}\left(\mathbf{b}_{4}\right)U\left(\mathbf{k-k}^{\prime}+\mathbf{b}_{1}-\mathbf{b}_{2}-\mathbf{b/2,b}\right).
\end{eqnarray}
Now we consider a long-range interaction, relevant for a FL near the Pomeranchuk instability.
In this case, the matrix element is non-negligible only if the first argument
of $U$ is as small as possible, which means that  $\mathbf{b}=2(\mathbf{b}_{2}-\mathbf{b}_{1})$ and $\bk\approx\bk'$:
\begin{eqnarray}
M_{\mathbf{kp\rightarrow k}^{\prime}\mathbf{p}^{\prime}}
= \sum_{{\bf b_1}\dots
\mathbf{b}_{4}}\delta_{\mathbf{k-k}^{\prime}\mathbf{+b}_{1}-\mathbf{b}_{2}\mathbf{,p}^{\prime}-\mathbf{p+b}_{3}-\mathbf{b}_{4}}u_{\mathbf{k}^{\prime}}^{\ast}\left(\mathbf{b}_{1}\right)u_{\mathbf{k}}\left(\mathbf{b}_{2}\right)u_{\mathbf{p}^{\prime}}^{\ast}\left(\mathbf{b}_{3}\right)u_{\mathbf{p}}\left(\mathbf{b}_{4}\right)U\left(\mathbf{k-k}^{\prime},2({\bf b}_2-{\bf b_1})\right).
\label{umklapp}
\end{eqnarray}
\end{widetext}
We see that the condition for an Umklapp process becomes very stringent:~\cite{maslov1}
since $\bk\approx \bk'$, the 
 momentum conservation condition $\bk-\bkp=\bpp-\bp+{\bar{\bf b}}$ can only be satisfied 
at special points, where $ \bpp-\bp\approx {\bar{\bf b}}$ and ${\bar{\bf b}}\equiv {\bf b}_4-{\bf b}_3+{\bf b}_1-{\bf b}_2$ is just another reciprocal lattice vector. 
As Fig.~\ref{fig:fig1}{\em a} shows, this is only possible if $\bp$ and $\bp'$ are located at the edges of the Brillouin zone (and the FS is open). The volumes (areas) around the special points are small--in proportion to a small momentum transfer ${\bar q}$. The corresponding scattering rate is smaller than the {\em transport} rate of {\em ee} collisions by a factor of ${\bar q}^D$. For HMM criticality, where ${\bar q}\propto T^{1/3}$, this implies that the contribution to the resistivity from the process depicted in Fig.~\ref{fig:fig1}{\em a} scales as $T^{2(D+1)/3}$, i.e., as $T^{8/3}$ in 3D and as $T^{2}$ in 2D. In both cases, the exponents are larger or equal than $2$. This means that the NFL contribution to the resistivity is smaller (3D) or comparable (2D) to the FL ($T^2$) contribution, arising from Umklapp scattering in the channels that are not affected by the proximity to a QCP, e.g., from the charge channel in the 
vicinity of a magnetic instability. 
In addition, processes in Fig.~\ref{fig:fig1}{\em a} are, in fact, "pseudo-Umklapps" because they can be viewed as normal processes on a closed (hole) FS.  The "real" Umklapps, shown in Fig.~\ref{fig:fig1}{\em b}, can occur only if the constraint of small momentum transfer is relaxed or else, near half-filling, when the \lq\lq gap\rq\rq\/ between the FS and the edges of the Brillouin zone is small. Half-filling, however, is more likely to result in a finite-$q$ instability of the ground state, e.g., antiferromagnetism, rather than in a Pomeranchuk QTP.
From now on, our analysis will be focused on normal {\em ee} collisions, 
the matrix element of which is given by Eq.~(\ref{umklapp}) with ${\bf b}_1-{\bf b}_2={\bf b}_3-{\bf b}_4$.
The corresponding scattering probability, averaged over spins of the initial states and summed over spins of the final states, reads
\begin{eqnarray}
W_{\mathbf{kp\rightarrow k}^{\prime}\mathbf{p}^{\prime}} & = &\frac{1}{4} \sum_{ij\gamma\delta}\left|\delta_{\alpha\gamma}\delta_{\beta\delta}M_{\mathbf{kp\rightarrow k}^{\prime}\mathbf{p}^{\prime}}-\delta_{\alpha\delta}\delta_{\beta\gamma}M_{\mathbf{kp\rightarrow p}^{\prime}\mathbf{k}^{\prime}}\right|^{2}\notag\\
 & = & \left|M_{\mathbf{kp\rightarrow k}^{\prime}\mathbf{p}^{\prime}}\right|^{2}+\left|M_{\mathbf{kp\rightarrow p}^{\prime}\mathbf{k}^{\prime}}\right|^{2}\notag\\
& &-\text{Re}\left(M_{\mathbf{kp\rightarrow k}^{\prime}\mathbf{p}^{\prime}}M^*_{\mathbf{kp\rightarrow p}^{\prime}\mathbf{k}^{\prime}}\right).
\end{eqnarray}
$W_{\mathbf{kp\rightarrow k}^{\prime}\mathbf{p}^{\prime}}$ has certain
symmetries. 
First, we assume the microreversibility property 
\begin{equation}
W_{\mathbf{kp\rightarrow k}^{\prime}\mathbf{p}^{\prime}}=W_{\mathbf{k}^{\prime}\mathbf{p}^{\prime}\mathbf{\rightarrow kp}}.\label{mr}
\end{equation}
In addition, since electrons are indistinguishable,
\begin{equation}
W_{\mathbf{k},\mathbf{p}\to \mathbf{k}^{\prime}\mathbf{p}^{\prime}}=W_{\mathbf{k}\mathbf{p}\to\mathbf{p}^{\prime}\mathbf{k}^{\prime}}=W_{\mathbf{p}\mathbf{k}\to\mathbf{k}^{\prime},\mathbf{p}^{\prime}}=W_{\mathbf{p}\mathbf{k}\to\mathbf{p}^{\prime}\mathbf{k}^{\prime}}.\label{ind}
\end{equation}
Finally, combining (\ref{mr}) and (\ref{ind}), we obtain 
\begin{equation}
W_{\mathbf{kp\rightarrow k}^{\prime}\mathbf{p}^{\prime}}=W_{\mathbf{k}^{\prime}\mathbf{p}^{\prime}\rightarrow\mathbf{kp}}=W_{\mathbf{p}^{\prime}\mathbf{k}^{\prime}\rightarrow\mathbf{pk}}
\end{equation}

With only normal {\em ee} collisions taken into account,  the total electron momentum is conserved, i.e.,
\begin{equation}
\int_\bk \bk I_{\mathrm{ee}} =0.\label{int0}
\end{equation}

Notice that although $I_{\mathrm{ei}}$ is written down in its most general form that holds true as long as $w_{\bk,\bkp}$
obeys unitarity~\cite{physkin,sturman84,comment_ei},  in writing down $I_{\mathrm{ee}}$ we have already assumed that $W_{\mathbf{k,p\rightarrow k}^{\prime}\mathbf{p}^{\prime}}$ obeys the microreversibility condition (\ref{mr}).

\subsection{General properties of the solution}
\label{sec:general}

Before proceeding with a more detailed analysis of the BE, we make a few general comments.
\begin{enumerate}
\item{{\bf \lq\lq Hidden\rq\rq\/ phonons.}}
The linearized form of the steady-state BE assumes implicitly
that the electron-phonon interaction is also present in the system;
otherwise, the total electron energy will increase indefinitely due
to the work done by the electric field. As usual (see, e.g., Ref.~\onlinecite{levinson}),
we assume that the temperature is low enough so that
one can neglect a direct electron-phonon contribution to the resistivity
(which requires that $\tau_{\mathrm{eph}}\gg\tau_{\mathrm{ee}},\tau_{\mathrm{ei}}$,
where $\tau$'s are the transport scattering times for corresponding
processes) but high enough so that, for a fixed electric
field, the electron-phonon interaction can still equalize the electron
and lattice temperatures (which requires that the work done by the electric
field on the energy relaxation length is much smaller than the temperature).
\item{{\bf Parity of a non-equilibrium part of the distribution function.}}
A linear in ${\bf E}$ term in $f_{\bk}$
can be written as 
\begin{equation}
\delta f_{\bk}\equiv f_{\bk}-n_{\bk}=-\mathbf{A}_{\mathbf{k}}\cdot\mathbf{E}Tn'_\bk,\label{dist}
\end{equation}
where ${\bf A}_{\bk}$ contains explicitly only the effects of the {\em
ee} and {\em ei} interactions. At low enough temperatures (as specified in the previous paragraph), the electron-phonon interaction
shows up only in the next--quadratic--term and does not affect the
resistivity directly. At even lower temperatures, when $\tau_{\mathrm{ee}}\gg\ti$, {\em ee} scattering
can be treated as a perturbation to {\em ei} scattering. In this case, ${\bf A}_\bk$ is determined by the crystal symmetry and
by the {\em ei} scattering probability and, without specifying both of them, no properties of  ${\bf A}_\bk$ can be further inferred. 
However, if the {\em ei} scattering probability satisfies the microreversibility condition,~\cite{levinson} i.e., $w_{\bk,\bk'}=w_{\bk',\bk}$, then ${\bf A}_\bk$  is odd in ${\bk}$.
Indeed, reversing the sign of $\bk$ in the BE and relabeling $\bk'\to-\bk'$, we obtain
\begin{equation}
e\bv_{\bk}\cdot {\bf E}n'_{\bk}=\int_{\bk'} 
w_{-\bk',-\bk}\left(f_{-\bk'}-f_{-\bk}\right)\delta\left(\ek-\ekp\right).
\end{equation}
Using time-reversal symmetry $w_{\mathbf{k,k}^{\prime}}=w_{\mathbf{-k}^{\prime},-\mathbf{k}}$
(which is guaranteed in the absence of the magnetic field and magnetic order) 
and microreversibility, we see that ${\bf A}_{-\bk}=-{\bf A}_{\bk}$.
This is the property of the non-equilibrium distribution function we will be
using later on. To simplify the presentation,
we will first use a model form of the {\em ei} collision integral, namely, a relaxation-time approximation $I_{\mathrm{ei}}=\left(f_{\bk}-n_{\bk}\right)/\tau_{\mathrm{i}},$ (cf. comment in Ref.~\onlinecite{comment_RTA}), which allows for a closed-form solution, and then extend the proof
for the general form of $f_\bk$ given by Eq.~(\ref{dist}).  

However, one has to keep in mind that, beyond the Born approximation, microreversibility is not
a general principle but a consequence of two microscopic symmetries, i.e.,
symmetries with respect to 
time-- and space-inversions, and is thus absent in non-centrosymmetric systems.~\cite{blokh}
\item
{\bf No disorder--no steady-state linear-response regime}. Since the momentum is conserved in normal collisions,
the collision integral (\ref{iee2}) is nullified by a combination $\mathbf{B\cdot k}$,
where $\text{\textbf{B}}$$ $ is $\bk$-independent but otherwise arbitrary. This means that there
is no unique steady-state solution in the linear-response regime. 
Obviously, the steady-state solution is absent because the total momentum of the electron system (per unit volume), ${\bf K}=\int_{\bk}{\bk}f_{\bk}$, increases with time.  Indeed, restoring the time and spatial derivatives in the BE, multiplying it by $\bk$ and integrating over $\bk$
we obtain
\beq
\frac{\partial K_{i}}{\partial t}+\frac{\partial \Pi_{ij}}{\partial x_{j}}=e\int_{\bk}k_{i}\frac{\partial f_\bk}{\partial k_{j}}E_{j},
\eeq
where $\Pi_{ij}=\int_{\bk}k_i v_j f_{\bk}$. Integrating by parts in the right-hand side
 and taking into account
that the number density $N=\int _{\bk}f_{\bk}$, we obtain 
\beq
\frac{\partial K_{i}}{\partial t}+\frac{\partial \Pi_{ij}}{\partial x_{j}}=-eNE_{i}
\eeq
The left-hand side is just the continuity equation while the right-hand side is the total force per unit volume.
Therefore, although the electron liquid is not, generally speaking, Galilean-invariant,  it is 
accelerated as a whole by the electric field. (In a crystal, an increase of the momentum in time leads to Bloch oscillations of the current; the current averaged over time is equal to zero.) Therefore, one needs to invoke impurity scattering in order to render the problem well-defined.~\cite{comment_compensated}

\item{{\bf  No lattice--no $T^2$ term in the resistivity.}} 
Adding just disorder but no lattice does not give rise to a $T^2$ term in the resistivity. 
Notice that this statement is weaker than \lq\lq the {\em ee} interaction does not effect the resistivity at all\rq\rq\/, which is true if $w_{\bk',\bk}$ depends only on the scattering angle but not on the electron energy. The simplest case is that of point-like impurities,
when $w_{\bk,\bk'}=1/\nu(\ek)\ti$, where $\nu(\ek)$  is the density of states (per one spin component) and $\ti$ is a constant. In this case,
 the BE reduces to
\begin{equation}
-e\bv_\bk\cdot {\bf E}n'_\bk=-\frac{f_\bk-{\bar f}}{\ti}-I_{\mathrm{ ee}},
\label{tra}
\end{equation}
where ${\bar f}$ is an average of $f_\bk$ the directions of $\bk$.
In the absence of lattice, $\bk=m\bv$ and hence the electric current ${\bf j}=-2e\int_{\bk}\bv f_\bk=-(2e/m)\int_\bk \bk f_\bk$.
Now one can multiply the BE equation by $\bv$ and integrate over $\bk$, upon which $I_{\mathrm{ee}}$--in accord with (\ref{int0})-- drops out, and obtain a relation between ${\bf j}$ and 
${\bf E}$ directly, without solving for $f_\bk$:
\begin{equation}
-e\int_{\bk}\bv_\bk (\bv_\bk\cdot{\bf E})n'_\bk=\frac{m}{2e\ti}{\bf j}
\end{equation}
The resuling conductivity $\sigma=ne^2\ti/m$ does not contain any effects of the {\em ee} interaction, except
for FL renormalizations of $m$ and $\ti$. 

The same is true if the scattering probability depends only on the angle between $\bk$ and $\bk'$. Parameterizing $w_{\bk,\bk'}$ as
\begin{equation}
w_{\bk,\bk'}={\bar w}\left(\ek,{\hat \bk}\cdot {\hat\bk'}\right)/\nu(\ek),
\end{equation}
we expand $f_{\bk}$ and ${\bar w}\left(\ek,{\hat \bk}\cdot {\hat\bk'}\right)$ over a complete basis set of, e.g., Legendre polynomials in 3D:
\begin{eqnarray}
f_{\bk}&=&\sum_{\ell}f^{\{\ell\}}(\ek){\mathcal P}_{\ell}(\cos\theta);\nonumber\\{\bar w}&=&\sum_{\ell}{\bar w}^{\{\ell\}}(\ek){\mathcal P}_{\ell}(\cos\theta){\mathcal P}_{\ell}(\cos\theta')+W_{o},
\end{eqnarray}
where $\theta$ ($\theta'$) is the angle between ${\bf E}$ and  $\bk$ ($\bk'$), and $W_{o}$ is an odd function of the polar angles that vanishes when substituted into the collision integral. In terms of angular harmonics, the BE reduces to
\begin{eqnarray}
-ev_{\bf k}\cdot {\bf E} n'_{\bk}=-\sum_{\ell}\frac{f^{\{\ell\}}}{\tau^{\{\ell\}}(\ek)}{\mathcal P}(\cos\ell\theta)-I_{\mathrm{ee}}
\label{BE_l}
\end{eqnarray}
with
\begin{equation}
\frac{1}{\ti^{\{\ell\}}(\ek)}\equiv {\bar w}^{\{0\}}(\ek)-\frac{1}{2\ell+1}{\bar w}^{\{\ell\}}(\ek).
\end{equation}
If $\ti^{\{\ell\}}(\ek)$ does not depend on the electron energy, one proceeds in the same way as for point-like  impurities, i.e., 
one obtains a direct relation between ${\bf j}$ and ${\bf E}$ by multiplying (\ref{BE_l}) by $\bv$ and integrating over $\bk$. The resulting (Drude) conductivity $\sigma=ne^2\ti^{{\mathrm tr}}/m$ contains the transport time 
$\ti^{{\mathrm tr}}\equiv \ti^{\{1\}}$ but no effects of the {\em ee} interaction (again, up to FL renormalizations).

If ${\bar w}$ does depend  on the electron energy, as it is often the case for semiconductors, the integral $\int_\bk \bk I_{\mathrm{ei}}$ does not reduce to the electric current, and one needs to solve for $f_\bk$  in order to find the conductivity.
Since the {\em ee} interaction affects $f_\bk$,  the conductivity is also affected. However, as we show in Sec.~\ref{sec:en_dep_tau}, the effect leads only to a $T^4$ term in the resistivity (or $T^4\ln T$ in 2D). 

\end{enumerate}

\section{\label{sec3} 
{Electron-electron contribution to the resistivity}}
\subsection{Do normal {\em ee} collisions affect the resistivity?}
It may seem that the reverse statement to the heading of item \#4 in the previous
Section (\lq\lq no lattice-no $T^2$ term in the resistivity\rq\rq\/) should be \lq\lq a $T^2$ term in the resistivity 
occurs in the presence of both disorder and lattice\lq\lq\/. Indeed, while disorder takes care of momentum
relaxation, lattice breaks the Galilean invariance. As a result,   $\bv_\bk=\partial\ek/\partial\bk\neq \bk/m$, which means that momentum
conservation does not imply current conservation, and one cannot obtain a relation between the current and the electric field
without actually solving the BE. In general, therefore, one should expect a $T^2$ term in the resistivity. While it is really the case in 3D,  it turns out that the conservation laws in 2D forbid the $T^2$ term for a convex and simply-connected  but otherwise arbitrary FS.

\subsection{Low temperatures: perturbation theory}
\label{sec:lowT}
In this Section, we discuss the case of low temperatures, when the {\em ee} collisions are less frequent than the {\em ei} ones. In this case, the {\em ee} contribution can be found via the perturbation theory with respect to $I_{\mathrm{ee}}$.
We begin with the simplest--isotropic-- model for electron-impurity scattering, 
when the BE is given by (\ref{tra}). However, we keep the dependence of the {\em ei} relaxation time on the electron energy for the time being.

At the first step, we solve (\ref{tra}) with $I_{\mathrm{ee}}=0$, which yields
\beq
g_{\mathbf{k}}^{\left(1\right)}=e\ti\left(\varepsilon_{\mathbf{k}}\right)\mathbf{v}_{\mathbf{k}}\mathbf{\cdot E/}T.
\eeq
Next, we substitute $g_{\mathbf{k}}^{\left(1\right)}$ back into (\ref{tra}) and find a correction due to $I_{\mathrm{ee}}$
\beq
g_{\mathbf{k}}^{\left(2\right)}=\frac{\ti\left(\varepsilon_{\mathbf{k}}\right)}{Tn_{\mathbf{k}}^{\prime}}I_{ee}\left[g_{\mathbf{k}}^{\left(1\right)}\right]
\eeq
The corresponding correction to the $ij$ component of the conductivity tensor is given by 
\begin{widetext}
\begin{eqnarray}
\delta\sigma_{ij} & = & -2\frac{e^{2}}{T}\int\frac{d^{D}q}{\left(2\pi\right)^{D}}\int\int\int d\omega d\varepsilon_{\mathbf{k}}d\varepsilon_{\mathbf{p}}\oint\oint\frac{da_{\mathbf{k}}}{v_{\mathbf{k}}}\frac{da_{\mathbf{p}}}{v_{\mathbf{p}}}W_{\mathbf{k,p}}\left(\mathbf{q},\omega\right)
\boldsymbol{\ell}_\bk^i\Delta\boldsymbol{\ell}^j\notag\\
& &\times n\left(\varepsilon_{\mathbf{k}}\right)n\left(\varepsilon_{\mathbf{p}}\right)\left[1-n\left(\varepsilon_{\mathbf{k}}-\omega\right)\right]\left[1-n\left(\varepsilon_{\mathbf{p}}+\omega\right)\right]
\delta\left(\varepsilon_{\mathbf{k}}-\varepsilon_{\mathbf{k-q}}-\omega\right)\delta\left(\varepsilon_{\mathbf{p}}-\varepsilon_{\mathbf{p}+\mathbf{q}}+\omega\right)\label{sigma_gen}
\end{eqnarray}
\end{widetext}
Here, $\bq\equiv \bk-\bk'=\bp'-\bp$ is the momentum transfer, $da_\bk$ is the surface (line) element of an isoenergetic surface (contour) at energy $\ek$ in 3D (2D), and $\Delta \boldsymbol{\ell}
\equiv\ti (\ek)\mathbf{v}_{\mathbf{k}}+\ti(\ep)\mathbf{v}_{\mathbf{p}}-\ti(\ek-\omega)\mathbf{v}_{\mathbf{k}-\mathbf{q}}-\ti(\ep+\omega)\mathbf{v}_{\mathbf{p}+\mathbf{q}}$ is a vector measuring the change in the total  \lq\lq vector mean free path \rq\rq\/ 
$\boldsymbol{\ell}_\bk\equiv \bv_{\bk}\ti(\ek)$ due to {\em ee} collisions.
The energy transfer was introduced by re-writing the energy conservation law as
$\delta\left(\varepsilon_{\mathbf{k}}+\varepsilon_{\mathbf{p}}-\varepsilon_{\mathbf{k}^{\prime}}-\varepsilon_{\mathbf{p}^{\prime}}\right)=\int d\omega\delta\left(\varepsilon_{\mathbf{k}}-\varepsilon_{\mathbf{k}^{\prime}}-\omega\right)\delta\left(\varepsilon_{\mathbf{p}}-\varepsilon_{\mathbf{p}^{\prime}}+\omega\right)$.
The scattering probability $W_{\bk,\bp}(\bq,\omega)\equiv W_{\bk,\bp\to\bk-\bq,\bp+\bq}$
is now allowed to depend on $\omega$.
Using the symmetry properties of $W_{\bk,\bp}(\bq,\omega)$, one can cast (\ref{sigma_gen}) into a more symmetric form

\begin{widetext} 
\begin{eqnarray}
\delta\sigma_{ij} & = & -\frac{e^{2}}{2T}\int\frac{d^{D}q}{\left(2\pi\right)^{D}}\int\int\int d\omega d\varepsilon_{\mathbf{k}}d\varepsilon_{\mathbf{p}}\oint\oint\frac{da_{\mathbf{k}}}{v_{\mathbf{k}}}\frac{da_{\mathbf{p}}}{v_{\mathbf{p}}}W_{\mathbf{k,p}}\left(\mathbf{q},\omega\right)  \Delta\boldsymbol{\ell}^i\Delta\boldsymbol{\ell}^j\notag\\
 &  & \times n\left(\varepsilon_{\mathbf{k}}\right)n_{\mathbf{p}}\left(\varepsilon_{\mathbf{p}}\right)\left[1-n\left(\varepsilon_{\mathbf{k}}-\omega\right)\right]\left[1-n\left(\varepsilon_{\mathbf{p}}+\omega\right)\right]
\delta\left(\varepsilon_{\mathbf{k}}-\varepsilon_{\mathbf{k-q}}-\omega\right)\delta\left(\varepsilon_{\mathbf{p}}-\varepsilon_{\mathbf{p}+\mathbf{q}}+\omega\right).
\label{sigma_gen_1}\end{eqnarray}
\end{widetext}
Now let us count the powers of $T$ in (\ref{sigma_gen_1}). Each of the three energy integrals (over $\omega$, $\ek$, and $\ep$) gives a factor of $T$ which, in a combination with the overall $1/T$ factor, already gives a $T^2$ dependence, as is to be expected for a FL.  The $T^2$ result holds as long as the
integral over $q$ does not introduce additional $T$ dependence. This is the case
in the FL regime, when typical $q$ are of order of the ultraviolet cutoff of the problem,
i.e., the smallest of the three quantities: the reciprocal lattice vector, a typical size of the FS, and the inverse radius of the {\em ee} interaction. In this case, the $\omega$ dependence of $W_{\mathbf{k,p}}\left(\mathbf{q},\omega\right)$ can be neglected.
The energy dependence of $\ti$ contributes only to higher order terms in $T$ and we neglect it for the time being as well, so that $\Delta\boldsymbol{\ell}=\ti\Delta\bv$
with \beq\Delta\bv=\bv_\bk+\bv_\bp-\bv_{\bk-\bq}-\bv_{\bp+\bq}\eeq being the change in the electron current due to {\em ee} collisions. Finally, since the integrals of the combinaton of the Fermi functions over energies already produce a factor of $T^2$, 
electrons can be projected onto the FS in the rest of the formula. This means that one can drop $\omega$ in both $\delta$-functions and perform the surface integrals over the FS. 

We pause here to remark that neglecting $\omega$ in the $\delta$-functions does {\em not} mean
performing an expansion in $\omega/\ek$, $\omega/\ep$, etc. In fact, all quasiparticles energies ($\ek$, $\varepsilon_{\bk-\bq}$, etc.) are equal to zero because the electrons were projected onto the FS! What it really means is that the $\delta$-functions impose constraints on the angles between $\bk$ and $\bq$ (and $\bp$ and $\bq$) with electrons' momenta being on the FS. Typical values of these angles are determined by the ratio of typical $q$ ($\equiv {\bar q}$) to $k_F$. In a system with a short-range interaction, ${\bar q}\sim \min\{k_F,1/a_0\}$, where $a_0$ is the lattice spacing; therefore, typical angles are of order unity. On the other hand, typical $\omega$ ($\equiv{\bar\omega}$) are of order $T$, and corrections to angles due to finite $\omega$ are small as long as $T\ll \min\{\varepsilon_F,W\}$, where $W$ is the bandwidth; the last condition is implied anyhow to be in the FL regime. If the interaction radius, $r_0$, is much longer than both the lattice spacing and the Fermi wavelength, ${\bar q}$ is small but in proportion to $1/r_0$ rather than to $T$, while ${\bar\omega}$ is still of order $T$. This means that effective ultraviolet energy scale is reduced to $v_F/r_0$, and the FL description is valid only at low energies, where the effect of a finite energy transfer on the kinematics of collisions is negligible. This can be illustrated for a simple example of the quadratic spectrum, when the angle between, e.g., $\bk$ and $\bq$, satisfies $\cos\theta_{\bk,\bq}=(q^2/2m+\omega)/v_Fq$. Neglecting $\omega$ is justified as long as $T\ll {\bar q}^2/2m$. Notice that this simplification is valid even near QPT (cf. Sec.~\ref{sec:pom}), where the scaling dimensions of $\omega$ and ${\bar q}$ are different: ${\bar\omega}\sim T$ but ${\bar q}\propto T^{1/3}$. A characteristic temperature, below which the condition ${\bar\omega}\sim T \ll{\bar q}^2/2m\propto T^{2/3}$ is satisfied, coincides with the scale below which the quasiparticle description breaks down, which is the regime of main interest for quantum phase transitions.  

As another remark, typical $q$ may be different for different observables.   What we said above is true for the {\em leading} term in the {\em ee} contribution to the electrical conductivity in all dimensions $D>1$, because the small $q$ behavior of the integrand in (\ref{sigma_gen_1}) is regularized by the $\Delta\boldsymbol{\ell}^i$ factors that vanish in the limit of $q\to 0$. (This is analogous to the regularizing effect of  the $1-\cos\theta$ factor in a transport cross-section for elastic scattering.)
However, when calculating the single-particle lifetime (the imaginary part of the self-energy)\cite{maslov04} and thermal conductivity\cite{lyakhov03} in 2D, one runs into infrared logarithmic divergences, which means that the infrared region of the momentum transfers ($q\sim T/v_F$) does contribute to the result. In those cases, neglecting $\omega$ in the $\delta$-functions is not justified. (The subleading
terms in the conductivity also require more care; see Sec.~\ref{sec:en_dep_tau} below.)

Coming back to the main theme, we focus now on the FL case, when one can also neglect $\omega$ in the scattering probability.
After all these simplifications,  the diagonal component of the conductivity reduces to 
\begin{widetext}
\begin{eqnarray}
\delta\sigma_{i i}  &= & -\frac{e^{2}}{2T}\ti^{2}\int\frac{d^{D}q}{\left(2\pi\right)^{D}}\int\int\int d\omega d\varepsilon_{\mathbf{k}}d\varepsilon_{\mathbf{p}}\oint\oint\frac{da^{F}_\bk}{v_\bk^F}\frac{da^{F}_\bp}{v_\bp^F}W_{\mathbf{k,p}}\left(\mathbf{q},0\right)
(\Delta\bv^i)^2
n\left(\varepsilon_{\mathbf{k}}\right)n\left(\varepsilon_{\mathbf{p}}\right)\left[1-n\left(\varepsilon_{\mathbf{k}}-\omega\right)\right]\left[1-n\left(\varepsilon_{\mathbf{p}}+\omega\right)\right]\notag\\
& &\times\delta\left(\varepsilon_{\mathbf{k}}-\varepsilon_{\mathbf{k-q}}\right)\vert_{\ek=0}\delta\left(\varepsilon_{\mathbf{p}}-\varepsilon_{\mathbf{p}+\mathbf{q}}\right)\vert_{\ep=0},
\end{eqnarray}
\end{widetext}
where superscript $F$ indicates that the corresponding quantity is evaluated at the FS.
Now the integrals over all energies can be performed with the help of  an
identity
\bea
&&\frac{1}{T}\int d\varepsilon_{1}\int d\varepsilon_{2}\int d\omega n\left(\varepsilon_{1}\right)n\left(\varepsilon_{2}\right)\left[1-n\left(\varepsilon_{1}-\omega\right)\right]\notag\\
&&\times\left[1-n\left(\varepsilon_{2}+\omega\right)\right]=\frac{2\pi^{2}}{3}T^{2},\label{ident}
\eea
 and we obtain a $T^2$ term in the conductivity with a prefactor given by a certain average
over the FS 
\bea
\delta\sigma_{ii}  &= & -\frac{\pi^2}{3}e^{2}\ti^{2}T^2\int\frac{d^{D}q}{\left(2\pi\right)^{D}}\oint\oint\frac{da^{F}_\bk}{v_\bk^F}\frac{da^{F}_\bp}{v_\bp^F}W_{\mathbf{k,p}}\left(\mathbf{q},0\right)
\notag\\
& &\times (\Delta\bv^i)^2\delta\left(\varepsilon_{\mathbf{k}}-\varepsilon_{\mathbf{k-q}}\right)\vert_{\ek=0}\delta\left(\varepsilon_{\mathbf{p}}-\varepsilon_{\mathbf{p}+\mathbf{q}}\right)\vert_{\ep=0}.\notag\\
\label{cond}\end{eqnarray}

Clearly, whether the leading correction to the residual conductivity
indeed scales as $T^{2}$ depends on whether the integral over the FS is nonzero.
Since the
integrand is positive, the integral may vanish only if $\Delta \bv=0$
under the energy conservation constraints imposed by the $\delta$-
functions. 
As a simple check, we apply Eq.~(\ref{cond}) for the Galilean-invariant case,
when $\bv_\bk=\bk/m$. In this case, $\Delta\bv=0$, as it should be.

We now consider a more general situation, when the $\omega$ dependence of the scattering probability is important, which is the case, e.g., near a QPT. In this case only two out of the three energy integrals can be performed explicitly and, instead of Eq.~(\ref{cond}), 
we obtain 
 \begin{eqnarray}
&&\delta\sigma _{ii } =-\frac{e^{2}\ti^2T^2}{2}\int \frac{d^{D}q}{\left( 2\pi
\right) ^{D}}\oint \oint \frac{da^F_{\bk}}{ v^F_{\bk} }%
\frac{da^F_{\bp}}{v^F_{\bp}}
R_{\mathbf{k,p}}
\left( \mathbf{q}\right) \notag\\
 &&\times
[ \Delta\bv^i]^2 \delta \left( \epsilon _{\mathbf{k}}-\epsilon _{\mathbf{k-q}}\right)\vert_{\epsilon_{\bk}=0}
 \delta \left( \epsilon _{
\mathbf{p}}-\epsilon _{\mathbf{p}+\mathbf{q}} \right)\vert_{\epsilon_{\bp}=0}
, 
\label{sigma_NFL}
\end{eqnarray}
 where 
 \beq
 R_{\mathbf{k,p}}\left(\mathbf{q}\right)\equiv \int d\omega\left(\omega^2/T^3\right)W_{\mathbf{k,p}}\left( 
\mathbf{q},\omega\right)N(\omega)
\left[N(\omega)+1\right]
\eeq
 and $N(\omega)$ is the Bose function.  
  For the effective interaction from Eq.~(\ref{hm1})  at the QPT  ($\delta=0$), power counting of (\ref{sigma_NFL}) gives
  $\delta\sigma_{ii}\propto T^{(D+2)/3}$, which coincides with the estimate based on the transport time (cf. Sec.~\ref{sec:pom}). As in the FL case, however, one needs to make sure that the prefactor is non-zero.
\subsection{Cases when the leading term vanishes}

\subsubsection{Isotropic system with an arbitrary spectrum}
\label{sec:isotropic}

The first case is that of an isotropic but otherwise arbitrary energy spectrum. 
Such a situation may arise due to relativistic effects. Another (pseudo-relativistic) example is weakly doped graphene with a negligibly small trigonal warping of the FS. 
Since $\varepsilon_\bk$
is a function of $|\mathbf{k}|$ only, the $\delta$-function constraints
in Eq.~(\ref{cond}) imply that 
$|\bk|=|\bk-\bq|$ and $|\bp|=|\bp+\bq|$.
Then, 
\begin{eqnarray}
v_{\bk}^{j} & = & 2\frac{\partial\varepsilon_{\bk}}{\partial(k^{2})}k_{j}=\xi(k) k_{j}\,;\notag\\
v_{\bk-\bq}^{j} & = & 2\left.\frac{\partial\varepsilon_{k}}{\partial(k^{2})}\right|_{
|\bk-\bq|=|\bk|}\!\!\!\!\!
\times(k_{j}-q_{j})
=\xi(k)(k_{j}-q_{j})\label{velocity-kq},\notag\\
\label{velocity-k}
\end{eqnarray}
where $\xi(k)\equiv v_\bk/k$.
Notice that the second line in Eq.~(\ref{velocity-k}) is not an expansion in small $q$ but
an exact relation. Substituting Eq.~(\ref{velocity-k}) 
(and similar expressions for $v_{\bp}^{j}$ and $v_{\mathbf{p}+\mathbf{q}}^{j}$)
into $\Delta \bv$, it is easy to see that $\Delta\bv$ vanishes
identically. 
Thus, there is no $T^{2}$ correction to the resistivity of a non-Galilean-invariant but isotropic
system. This result also holds for a general quadratic spectrum $\varepsilon_\bk=k_ik_j/2m_{ij}$, in which
case $v_j=k_i/m_{ji}$ and $\Delta\bv^j=0$.


Notice that, in contrast to the Galilean-invariant case (with $\ek=k^2/2m-\epsilon_F$),
when not only the $T^2$ term but all higher order terms
are absent, higher order ($T^4$, etc.) terms are non-zero for a non-parabolic spectrum.

\begin{figure}
\includegraphics[width=0.45\textwidth]{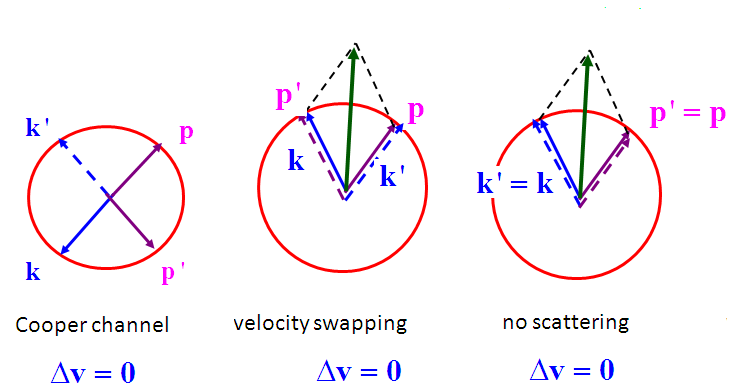} \caption{Isotropic case in 2D: three possible scattering processes none of which leads to current relaxation.}
\label{isotropic}
\end{figure}

\subsubsection{Approximate integrability: convex and simply connected Fermi surface in 2D}
\label{sec:convex}
\paragraph{{\bf Kinematics of {\em ee} collisions on a circular FS.}}
The fact that $T^2$ term in the resistivity is absent for an isotropic FS  does not mean that it is necessarily present for an anisotropic FS.  In fact, the $T^2$ term is also absent for a simply-connected and convex but otherwise arbitrary FS in 2D.~\cite{gurzhi_ee,maebashi97_98,maslov1}. Before considering the general case, however, let us study the simplest example of such a FS, i.e., a 2D circular FS with quadratic spectrum. Since this is just a Galilean-invariant case, we already know (cf. Sec.~\ref{sec:general}) that the {\em ee} interaction has no effect on the resistivity.  However, it  is instructive to see in a geometrical way how the $T^2$ term vanishes--this will be useful for the subsequent analysis of the general case in 2D. Geometrically, one needs to find two initial momenta, $\bk$ and $\bp$, belonging to the FS, such that the final momenta, $\bk-\bq$ and $\bp+\bq$, also belong to the FS.  
As shown in Fig. ~\ref{isotropic}, only three situations are possible:~\cite{gurzhi95} 1) Cooper channel, when the total initial and, therefore, the total final momentum is equal to zero; 2) swapping of velocities, when the initial momentum of one the electrons coincides with the final momentum of another electron and {\em vice versa}, i.e., $\bp=\bk-\bq$; 3) no scattering- this is the trivial case where the initial and final momenta of individual electrons are the same. For all of these cases, $\Delta\bv=0$ and thus the $T^2$ term is absent.
To see that these situations indeed exhaust all the possibilities, one can solve the momentum and energy conservation equations, 
i.e., $\mathbf{k}-\mathbf{k}^{\prime}=\mathbf{p}^{\prime}-\mathbf{p}=\mathbf{q}$ and $k^2=k^{\prime 2}; p^2=p^{\prime 2}$, subject to the additional constraint $k=p=k'=p'=k_F$. 
This leads to two equations: $q^2-2kq\mathrm{cos}\theta_{kq}=0$ and $q^2+2pq\mathrm{cos}\theta_{pq}=0$, where $\theta_{ij}$ denotes the angle between the vectors $\mathbf{i}$ and $\mathbf{j}$. The three possible solutions are: $\theta_{kq}-\theta_{pq}=\pi$, corresponding to case 1); $\theta_{kq}+\theta_{pq}=\pi$, corresponding to case 2); and $q=0$ for arbitrary $\theta_{kq}$ and $\theta_{pq}$ , corresponding to case 3). 

\begin{figure}
\includegraphics[width=0.4\textwidth]{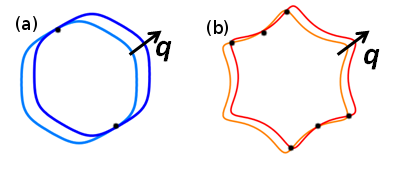} \caption{(a) A convex contour has at most two self-intersection points (marked by dots). (b) A concave contour can have more than two self-intersection points (six in the example shown).}
\label{intersection}
\end{figure}

\paragraph{{\bf Kinematics of {\em ee} collisions on a generic convex FS.}}
The situation described above is not specific to a circular FS in 2D but  occurs also for a generic convex FS, see Fig.~\ref{intersection}a.   Indeed, introducing a new variable $\mathbf{\bar{p}=-p}$ in (\ref{cond}) and using the time-reveral symmetry ($\varepsilon_{-\mathbf{p}}=\varepsilon_{\mathbf{p}}$)
and symmetries of the scattering probability, we obtain 
\begin{widetext} 
\begin{equation}
\delta\sigma_{ii}=-\frac{\pi^{2}}{3}e^{2}T^{2}\ti^{2}\int\frac{d^{2}q}{\left(2\pi\right)^{2}}\oint\oint\frac{da^F_{\mathbf{k}}}{v_{\mathbf{k}}}\frac{da^F_{\mathbf{\bar{p}}}}{v_{\mathbf{\bar{p}}}}W_{\mathbf{k,{\bar p}}}\left(\mathbf{q,0}\right)\left[\mathbf{v}_{\mathbf{k}}^{i}-\mathbf{v}_{\mathbf{\bar{p}}}^{i}-\mathbf{v}_{\mathbf{k}-\mathbf{q}}^{i}+\mathbf{v}_{\mathbf{\bar{p}}-\mathbf{q}}^{i}\right]^{2}\delta\left(\varepsilon_{\mathbf{k}}-\varepsilon_{\mathbf{k-q}}\right)\delta\left(\varepsilon_{\mathbf{\bar{p}}}-\varepsilon_{\mathbf{\bar{p}}-\mathbf{q}}\right).
\label{delsigma}
\end{equation}
\end{widetext}
For given $\mathbf{q}$, we must find two momenta satisfying
the relations $\varepsilon_{\mathbf{k}}=\varepsilon_{\mathbf{k-q}}$
and $\varepsilon_{\mathbf{\bar{p}}}=\varepsilon_{\mathbf{\bar{p}}-\mathbf{q}}$.
Geometrically, finding the solution to these two equations is equivalent to shifting the FS by $\mathbf{q}$, and finding the points of intersection between the original and the shifted FSs.
 A convex FS has at most two self-intersection points.  Therefore,
the equation $\varepsilon_{\mathbf{k}}=\varepsilon_{\mathbf{k-q}}$
has only two solutions. In addition, if $\mathbf{k}$ is a solution, then $-\mathbf{k}+\mathbf{q}$ is also a solution so that the roots
of the first equation form a set $\{\bk,-\bk+\bq\}$.
Since the second equation is the same, its two roots $\{{\bar \bp},-{\bar\bp}+\bq\}=\{-\bp,\bp+\bq\}$ must coincide with the roots of the first equation. This can happen  if 1) $\bk=-\bp$, which gives the Cooper channel or  if 2) $\bk=\bp+\bq$ which gives swapping. The situation with $q=0$, when no scattering occurs, is trivially possible.
For all  the scattering processes listed above, $\Delta \bv=0$ and the $T^2$ term vanishes. 

Being purely geometrical, the preceding analysis is equally valid for the NFL case, with the conclusion that the $T^{4/3}$ term vanishes as well.

\paragraph{{\bf Beyond the relaxation-time approximation.}}
Although the analysis above was based on Eq. ~(\ref{delsigma}), obtained in the relaxation-time approximation for \emph{ei} scattering, it can be readily extended for the general form of the {\em ei} collision integral  in Eq.~(\ref{iei}). 
The non-equilibirum part of the distribution function
in the presence of {\em ei} scattering only  is given by (\ref{dist}), which implies that  $g_{\mathbf{k}}^{(1)}$ in (\ref{dist1}) is replaced by 
$g_{\mathbf{k}}^{(1)}=
\mathbf{A}_{\mathbf{k}}\cdot \mathbf{E}$. The lowest-order iteration in {\em ee} scattering is to be found from an integral equation
\beq
I_{\mathrm{ei}}[g_{\mathbf{k}}^{(2)}]=\frac{1}{Tn'_\bk}I_{\mathrm{ee}}[{\bf A}_\bk\cdot{\bf E}n'_{\bk}].
\label{inteq}
\eeq
The {\em ei} collision integral can be viewed as an integral operator, the inverse of which is defined by
\beq
{\hat I}^{-1}_{{\mathrm ei}}[f_\bk]\equiv \sum_{\hat \bk} O_{\bk,\bk'}f_{\bk'},
\eeq
where ${\hat\bk}\equiv \bk/k$. Thanks to microreversibility,  $O_{\bk,\bk'}=O_{\bk',\bk}$.
A formal solution of (\ref{inteq}) is
\beq
g^{(2)}_\bk=\frac{1}{Tn'_{\bk}}{\hat I}_{\mathrm{ei}}^{-1}I_{\mathrm{ee}}[{\bf A}_\bk\cdot{\bf E}].
\eeq
Using the microreversibility property of $O_{\bk,\bk'}$ and the fact that $A_{\bk}$ is odd in $\bk$, it is easy to see that $g^{(2)}_\bk$ is odd in $\bk$ as well. 
This is all one really needs to repeat the steps of the previous analysis.
The correction to the conductivity
now contains a combination $\Delta \bv^{i}\Delta {\bf A}^{j}$, where $\Delta{\bf A}\equiv {\bf A}_\bk+{\bf A}_\bp-{\bf A}_{\bkp}-{\bf A}_{\bpp}$.  
Being odd in all momenta, $\Delta {\bf A}$ behaves in the same way as $\Delta\bv$ upon the change $\bp\to-\bp$. The scattering processes are classified in the same way as before,
and the vanishing of the $T^2$ term follows from the vanishing of $\Delta\bv$.

\paragraph{{\bf Approximate integrability.}}
A limited number of possible outcomes of the {\em ee} collisions means that our 2D system
behaves similar to a 1D system, where binary collisions do not lead to relaxation.
The analogy works because, to find the leading ($T^2$) term in the conductivity, it suffices to
project  electrons onto the FS, which is a line in 2D. Therefore, kinematics effectively becomes 1D and, although this is a 2D case, we have an integrable system. However, this analogy has certain limitations. First, the 2D case is integrable only with respect to charge but not thermal current relaxation, whereas there is no relaxation of all physical quantities in 1D.   Second, even the charge current relaxation is absent only up to next-order-terms in $T/\varepsilon_F$ (see Sec.~\ref{sec:T4}).
Third, not any FS line in 2D is integrable: concave and multiply-connected contours behave in a non-integrable way. With all these limitations in mind, we will refer to the 2D convex case as to "approximate integrability".

\subsubsection{Subleading corrections to the resistivity when the leading term is absent.}
\paragraph{{\bf Higher order term from electrons away from the FS}.}
\label{sec:T4}
To find the subleading correction for the case considered in the previous section, 
we go back to Eq.~(\ref{sigma_gen_1}), replace again $\ti$ by a constant in the scattering probability, but now, instead of neglecting $\omega$ in the $\delta$
functions, expand the product of the $\delta$ functions to second
order in $\omega$. The zeroth-order term, $\delta\left(\varepsilon_{\mathbf{k}}-\varepsilon_{\mathbf{k-q}}\right)\delta\left(\varepsilon_{\mathbf{p}}-\varepsilon_{\mathbf{p}+\mathbf{q}}\right)$, 
nullifies $\Delta\mathbf{v}^{i}$.  The odd in $\omega$ terms
vanish upon integration over $\varepsilon_{\bk,}\varepsilon_{\bp,}$ and
$\omega.$ In the FL case, this gives
\begin{widetext} 
\begin{eqnarray}
\delta\sigma_{ii} & = & \frac{1}{2}\frac{e^{2}}{T}\ti^{2}\int\frac{d^{D}q}{\left(2\pi\right)^{D}}\int\int\int d\omega\omega^{2}d\varepsilon_{\mathbf{k}}d\varepsilon_{\mathbf{p}}\oint\oint\frac{da_{\mathbf{k}}}{v_{\mathbf{k}}}\frac{da_{\mathbf{p}}}{v_{\mathbf{p}}}W_{\mathbf{k,p}}\left(\mathbf{q},0\right)\notag\\
 &  & \times\left[\Delta\mathbf{v}^{i}\right]^{2}n\left(\varepsilon_{\mathbf{k}}\right)n\left(\varepsilon_{\mathbf{p}}\right)\left[1-n\left(\varepsilon_{\mathbf{k}}-\omega\right)\right]\left[1-n\left(\varepsilon_{\mathbf{p}}+\omega\right)\right]\notag\\
 &  & \times\left[\delta^{\prime}\left(\varepsilon_{\mathbf{k}}-\varepsilon_{\mathbf{k-q}}\right)\delta^{\prime}\left(\varepsilon_{\mathbf{p}}-\varepsilon_{\mathbf{p}+\mathbf{q}}\right)-\frac{1}{2}\left\{ \delta^{\prime\prime}\left(\varepsilon_{\mathbf{k}}-\varepsilon_{\mathbf{k-q}}\right)\delta\left(\varepsilon_{\mathbf{p}}-\varepsilon_{\mathbf{p}+\mathbf{q}}\right)+\delta\left(\varepsilon_{\mathbf{k}}-\varepsilon_{\mathbf{k-q}}\right)\delta^{\prime\prime}\left(\varepsilon_{\mathbf{p}}-\varepsilon_{\mathbf{p}+\mathbf{q}}\right)\right\} \right].
\end{eqnarray}
\end{widetext} The derivatives of the $\delta$-functions produce the
same roots for $\mathbf{k}$ and $\mathbf{p}$ as the $\delta$-functions
themselves. However, integrating by parts, we make the derivatives
to act on $\left[\Delta\mathbf{v}^{i}\right]^{2}.$ Although
$\left[\Delta\mathbf{v}^{i}\right]^{2}$ vanishes for $\bk$ and $\bp$ satisfying energy and momentum
conservations,
its derivatives do not. This makes the integral non-zero. Since we now have
two more factors of $\omega$ the correction to the conductivity
scales as
 \beq
 \delta\sigma_{ii}\propto  T^{4}.
 \eeq
  In more detail, let $\mathbf{k}_{0}$ be one of
the roots of the equation $\varepsilon_{\mathbf{k}}=\varepsilon_{\mathbf{k-q}}.$
The corresponding root for $\mathbf{p}$ is then $\mathbf{p}_{0}=\mathbf{k}_{0}-\mathbf{q.}$
Expanding $\Delta\bv$ around the roots gives
\begin{eqnarray}
\Delta\mathbf{v}^{i} =
  \left(\left[\delta\mathbf{k-}\delta\mathbf{p}\right]\cdot\nabla\right)\left(\mathbf{v}_{\mathbf{k}_{0}}^{i}-\mathbf{v}_{\mathbf{k}_{0}-\mathbf{q}}^{i}\right),
\end{eqnarray}
where $\delta\mathbf{k}\equiv \bk-\bk_{0}$ and $\delta\bp\equiv \bp-\bk_0+\bq$.
Subsequent integration proceeds as in the integral 
\beq
\int\int dxdy\delta^{\prime}\left(x\right)\delta^{\prime}\left(y\right)\frac{1}{2}\left(x-y\right)^{2}=-\int dx\delta^{\prime}\left(x\right)x=1,
\eeq
where $\delta\mathbf{k}$ and $\delta\mathbf{p}$ play the roles of $x$
and $y$ (and similarly for an integral with a product $\delta''(\dots)\delta(\dots)$).
 Further cancelations for a particular FS may make the $T$
dependence even weaker but the generic answer is $T^{4}.$

Clearly, going away from the FS produces an extra factor of $T^2$. Since $\omega\sim T$ in the NFL as well, the result for the NFL regime is obtained by multiplying the \lq\lq naive\rq\rq\/ estimate $\delta\sigma_{ii}\propto T^{4/3}$ by $T^2$, which gives $\delta\sigma_{ii}\propto T^{10/3}$. This is obviously subleading to the $T^2$ term resulting from the FL interaction in non-critical channels.

\paragraph{{\bf Energy-dependent electron-impurity relaxation time.}}
\label{sec:en_dep_tau}
In addition to the mechanism described above, there are other sources of higher than $T^2$ corrections to the conductivity;
one of them is the energy dependence of $\ti$ which we have neglected so far.
This mechanism operates even in a Galilean-invariant system: although {\em ee} collisions conserve the momentum, they redistribute electrons in the energy space and thus affect the conductivity, if $\ti$ depends on the energy.~\cite{debye54,levinson}
To estimate the magnitude of this effect, we apply Eq.~(\ref{sigma_gen_1}) to the Galilean-invariant case ($\bv=\bk/m$) and expand the impurity relaxation times entering the \lq\lq vector mean free path\rq\rq\/ as
$\ti\left(\varepsilon_{{\bf l}}\right)=\ti(0)+\ti^{\prime}\varepsilon_{{\bf l}}$, where $\ti'\equiv \partial\ti(\varepsilon_{{\bf l}})/\partial\varepsilon_{{\bf l}}|_{\varepsilon_{{\bf l}}=0}$. This yields 
\begin{eqnarray}
 \Delta\boldsymbol{\ell}
  = \frac{\ti^{\prime}\omega}{m}\left(\mathbf{k-p}-2\mathbf{q}\right).
  \label{kpq}
\end{eqnarray}
Since (\ref{sigma_gen})  contains two factors of $\Delta\boldsymbol{\ell}$,  and each of them is proportional
to $\omega$, we have an extra $\omega^2$ factor in the integrand. In 3D, this immediately gives a $T^4$ term
\beq
\delta\sigma^{ii}_{3D}\propto (\ti')^2T^4.
\eeq
In 2D, the situation is more delicate because the part of the integrand associated with the $\bk-\bp$ term in (\ref{kpq})
is logarithmically divergent. This a well-known \lq\lq 2D log singularity\rq\rq\/
that occurs, on a more general level, as the mass-shell singularity of the self-energy (see Ref.~\onlinecite{maslov04} and references therein). This is also the same singularity that one encounters when calculating the thermal conductivity in 2D (in the absence of impurity scattering).~\cite{lyakhov03} Indeed, our problem bears a formal similarity to that of the thermal conductivity because
the change in the thermal current ${\bf j}^T_\bk=\bv_\bk\ek$ due to {\em ee} collisions
 \bea
 {\bf j}^T_{\bk}+{\bf j}^T_{\bp}-{\bf j}^T_{\bk-\bq}-{\bf j}^T_{\bp+\bq}=\frac{
 \left(\bk-\bp-2\bq\right)\omega+\bq(\ek-\ep)}
  {m}
 \notag\\
  \eea
  contains the same term as $\Delta\boldsymbol{\ell}$ in (\ref{kpq}). The singularity can be resolved by the same method
  as in Ref.~\onlinecite{lyakhov03}. i.e., by considering a dynamically screened Coulomb interaction. The result is that, similar to the thermal conductivity, the conductivity contains an extra log factor as compared to the 3D case: 
  \beq
\sigma^{ii}_{2D}\propto (\ti')^2T^4\ln\left(\varepsilon_F/T\right).
\eeq
 The \lq\lq 2D log\rq\rq\/ does not occur in the $T^2$ term in the conductivity, if the latter is finite
due to broken integrability, which is the subject of the next section.

An extension to the NFL case is again, trivial, and we will not repeat the argument here.

\subsection{Non-integrable cases
}
\paragraph{{\bf Concave FS in 2D}.}
It follows from the previous discussion that whether the $T^2$ term is absent or present depends entirely on 
the FS having two or more than two self-intersection points. A concave FS in 2D can have more than two self-intersection points (cf. Fig. ~\ref{intersection}b), therefore there are more than two solutions for the initial momenta for given $\bq$. Some of these solutions still correspond to "integrable" processes, encountered already for a convex FS, but the remaining ones do relax the current. Therefore, a $T^2$ term survives in this case.

\begin{figure}
\includegraphics[width=0.4\textwidth]{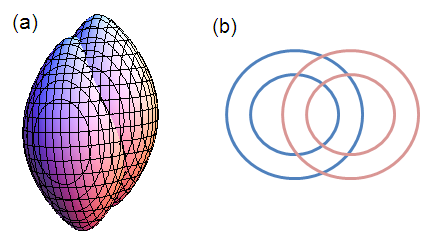} \caption{(a) A 3D FS has an infinite number of self-intersection points (a line). (b) A multiply connected FS has more than two self-intersection points.}
\label{dim_topo}
\end{figure}

\paragraph{{\bf 3D FS}.}
 In 3D, the manifold of intersection between the original and shifted FSs is  a line, see Fig. ~\ref{dim_topo}(a). 
 Therefore, the equation $\varepsilon_{\mathbf{k}}=\varepsilon_{\mathbf{{k}-{q}}}$ has infinitely many roots.
There is no correlation between
the roots of the equations $\varepsilon_{\mathbf{k}}=\varepsilon_{\mathbf{{k}-{q}}}$
and $\varepsilon_{\mathbf{p}}=\varepsilon_{\mathbf{{p}+{q}}}$. Geometrically,
this means that the initial momenta, $\mathbf{k}$ and $\mathbf{p}$, do not have 
to be in the same plane as the final ones,  $\mathbf{k}^{\prime}$ and
$\mathbf{p}^{\prime}$.
Therefore, an anisotropic (but not quadratic) FS in 3D allows for
a $T^{2}$ correction to  the resistivity.

The $T^{5/3}$ term in the NFL regime survives for the same reason as well. Therefore, our theory at least does not contradict
the experiments \cite{exp_fm} where such term was observed.

\paragraph{{\bf Multiply connected FS}.}
If the FS is multiply connected, a $T^2$ term in the resistivity is present, 
even if the individual FS sheets do not allow for a $T^2$ term on their own. Even more so, the individual sheets can even be isotropic.
 The reason is obvious from Fig.~\ref{dim_topo}(b) which shows an example of two circular FSs in 2D.
Clearly, the equation $\varepsilon_{\mathbf{k}}=\varepsilon_{\mathbf{{k}-{q}}}$
has more than two roots even in this case. Thus, according to our previous
arguments, there is no general reason for the vanishing of the $T^{2}$
term in such a situation. In Sec.~\ref{sec4} and Appendix \ref{app:twoband}, we discuss the two-band case in 2D in more detail.

\subsection{Weakly-integrable cases}
In this section, we consider two situations when integrability is broken only weakly.

\subsubsection{Quasi-2D metal}
The first case is a layered metal with a quasi-2D spectrum which,
for simplicity, we assume to be 
separable into the in- and out-of-plane parts as
\beq
\varepsilon_{\mathbf{k}}=\varepsilon^{||}_{\bk_{||}}
+\varepsilon_{k_{z}}^{z},
\label{q2d}
\eeq
where $\bk_{||}$ and $k_z$ are the in-plane and out-of-plane components of the momentum, 
correspondingly. 
In the tight-binding model with nearest-neighbor hopping, $\varepsilon_{k_{z}}^{z}=t_{\perp}\left[1-\cos\left(k_{z}c\right)\right]$,
where $c$ is the lattice spacing in the $z$-direction. The metal is in a quasi-2D regime when $t_\perp\ll\varepsilon_F$.
In regard to the in-plane part of the spectrum, $\varepsilon^{||}_{\bk_{||}}$, we assume that the corresponding energy 
contours are anisotropic
but convex so that, in the absence of inter-plane hopping, the $T^2$-term in the in-plane conductivity would be 
absent.
(If the planes are assumed to be Galilean-invariant, i.e.,  $\varepsilon^{||}_{\bk_{||}}=k_{||}^2/2m_{||}$, as in a \lq\lq corrugated cylinder model\rq\rq\/,
the $T^2$-term is trivially zero because 
the in- and out-of-plane components of the momentum are conserved independently, and
hence
$\bv^{||}_{\bk_{||}}+\bv^{||}_{\bp_{||}}-\bv^{||}_{\bk'_{||}}-\bv^{||}_{\bp'_{||}}=0$.)
To find the $T^2$-term in the in-plane conductivity, we use a method similar to that in Sec.~\ref{sec:T4}, i.e., we expand the
$\delta$-functions, except for that now we expand both in $\omega$ and $\varepsilon^{z}_{k_z}$. 
As we explained in Sec.~\ref{sec:lowT}, the expansion in $\omega$ is really an expansion in $\omega$ normalized 
by the appropriate ultraviolet energy scale of the problem. Likewise, the expansion in $\varepsilon^{z}_{k_z}$
is really an expansion in $t_{\perp}/\epsilon_{F}$, which is a natural small parameter for a quasi-2D system.
The zeroth-order term ($\omega=0$, $\varepsilon^{z}_{k_z}=0$) nullifies $\Delta\bv^{||}$. 
The first-order terms also vanish: the ones, proportional to $\omega$, do so by parity,   and the ones, proportional to  $\varepsilon^{z}_{k_z}$, do so
because the first-order derivatives of the $\delta$-functions nullify $(\Delta\bv^{||})^2$ after a single integration by parts.
Finally, the cross products in second-order terms, being odd in $\omega$, also vanish. 
Therefore, the only surviving second-order term is
\begin{widetext} 
\begin{eqnarray}
 &  & \delta\left(\varepsilon_{\mathbf{k}_{||}-\mathbf{q}_{||}}^{||}-\varepsilon_{\mathbf{k}_{||}}^{||}+\varepsilon_{k_{z}-q_{z}}^{z}-\varepsilon_{k_{z}}^{z}-\omega\right)\delta\left(\varepsilon_{\mathbf{p}_{||}+\mathbf{q}_{||}}^{||}-\varepsilon_{\mathbf{p}_{||}}^{||}+\varepsilon_{p_{z}+q_{z}}^{z}-\varepsilon_{p_{z}}^{z}+\omega\right)\notag\\
 & = &\frac{1}{2}\left[\left(\varepsilon_{k_{z}-q_{z}}^{z}-\varepsilon_{k_{z}}^{z}\right)^2+\omega^{2}\right]\delta^{{\prime\prime}}
 \left(\varepsilon_{\mathbf{k}_{||}-\mathbf{q}_{||}}^{||}-\varepsilon_{\mathbf{k}_{||}}^{||}\right)\delta\left(\varepsilon_{\mathbf{p}_{||}+\mathbf{q}_{||}}^{||}-\varepsilon_{\mathbf{p}_{||}}^{||}\right)+\frac{1}{2}\left[\left(\varepsilon_{p_{z}+q_{z}}^{z}-\varepsilon_{p_{z}}^{z}\right)^2+\omega^{2}\right]\delta\left(\varepsilon_{\mathbf{k}_{||}-\mathbf{q}_{||}}^{||}-\varepsilon_{\mathbf{k}_{||}}^{||}\right)
 \notag\\
 &&\times\delta^{{\prime\prime}}\left(\varepsilon_{\mathbf{p}_{||}+\mathbf{q}_{||}}^{||}-\varepsilon_{\mathbf{p}_{||}}^{||}\right)
 +\left[\left(\varepsilon_{k_{z}-q_{z}}^{z}-\varepsilon_{k_{z}}^{z}\right)\left(\varepsilon_{p_{z}+q_{z}}^{z}-\varepsilon_{p_{z}}^{z}\right)-\omega^2\right]\delta'\left(\varepsilon_{\mathbf{k}_{||}-\mathbf{q}_{||}}^{||}-\varepsilon_{\mathbf{k}_{||}}^{||}\right)\delta'\left(\varepsilon_{\mathbf{p}_{||}+\mathbf{q}_{||}}^{||}-\varepsilon_{\mathbf{p}_{||}}^{||}\right).
 \label{exp_omega_tperp}
\end{eqnarray}
\end{widetext} 
Equation (\ref{exp_omega_tperp}) contains two independent corrections. All terms proportional to $\omega^2$ produce a $T^4$
correction to the conductivity that exists even in a purely 2D system. All terms containing the squares of the out-of-plane dispersions produce a $T^2$ correction.~\cite{gurzhi_ee} Therefore,
\beq
\delta\sigma_{ii}=A_4T^{4}+A_2t_{\perp}^{2}T^{2},
\label{q2d}
\eeq
where $i=x,y$,
and constants $A_4$ and $A_2$ depend on details of the in-plane spectrum;  generically, $A_4\sim A_2$.
Equation~(\ref{q2d}) describes a dimensional crossover from the 2D-like regime ($\delta\sigma_{ii}\propto T^4$) at $T\gg t_{\perp}$ to the 3D-like regime ($\delta\sigma_{ii}\propto T^2$) for $T\ll t_{\perp})$.  Notice
that, in the 3D regime, the $T^2$-term in the {\em in-plane} conductivity depends on the {\em out-of-plane} hopping.

In the NFL regime, Eq.~(\ref{q2d}) is replaced by
\beq
\delta\sigma_{ii}=A_4T^{10/3}+A_2t_{\perp}^{2}T^{4/3}.
\label{q2d_NFL}
\eeq

\subsubsection{Conductivity near the convex-concave transition}
\label{sec:bi2te3}
In this Section, we consider a FS near a convex-concave transition which occurs  when  the Fermi energy
goes above a certain  threshold value $\varepsilon_c$.~\cite{pal} Such a situation is encountered, e.g., 
 in the case of surface states of the Bi$_{2}$Te$_{3}$ family
of 3D topological insulators, where the electron spectrum can be approximated by\cite{fu} $
\epsilon_{\mathbf{k}}^{\pm}=\pm\sqrt{v_F^2k^2+\lambda^2 k^6\mathrm{cos}^2(3\theta)}$,
with  $\theta$ being  the polar angle.
Corresponding isoenergetic contours are shown in Fig. \ref{fig3}(a). We will be interested in the vicinity of the convex/concave transition, when $|\varepsilon_F-\varepsilon_c|\equiv |\Delta|\ll \varepsilon_c$.

\begin{figure}
\includegraphics[width=0.4\textwidth]{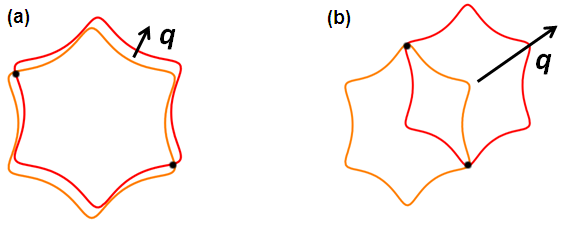} \caption{(color online). (a) Even in the concave case, there can be only two intersection points if $\mathbf{q}$
is not along a special direction. (b) For $q$ larger than a critical
value, there are only two intersection points.}
\label{fig2} 
\end{figure}

\begin{figure}
\includegraphics[width=0.5\textwidth]{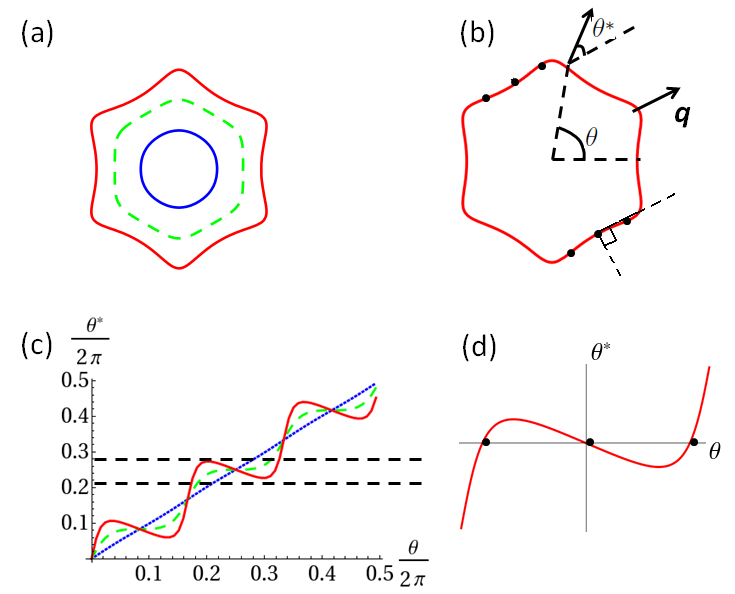} \caption{(color online). (a) Isoenergetic contours 
for the surface states of the  Bi$_{2}$Te$_{3}$ family
of 3D topological insulators. The dashed  line corresponds to the critical energy for the convex-concave transition. (b) For small $q$, points (black dots), where the normal to the
FS is perpendicular to $\mathbf{q}$, are the points of self-intersection. 
(c) $\theta^{*}$ vs $\theta$ {[}as defined in panel (b){]}. Solid:
$\varepsilon_F>\varepsilon_{c}$; dashed: $\varepsilon_F=\varepsilon_{c}$; dotted:
$\varepsilon_F<\varepsilon_{c}$;. (d) A zoom of the non-monotonic part of
the graph in panel (c) 
Reproduced from Ref.~\onlinecite{pal}, courtesy of the APS.}

\label{fig3} 
\end{figure}

We first consider the case of $\Delta\gg T$, when the isoenergetic
contours near the Fermi energy are concave and thermal population of concave isoenergetic contours can be neglected. Obviously, $\Delta\mathbf{v}$ in Eq.~(\ref{delsigma}) shows a critical behavior: it is zero on the convex
side and non-zero on the concave side of the transition. However, there are two other
quantities which also show a critical behavior. As
Figs.~\ref{fig2}(a) and \ref{fig2}(b) illustrate, even a concave FS
does not necessarily have more than two self-intersection points: this happens only if the FS is shifted along certain directions  that lie close to high symmetry axes, i.e., $\bq$ lies within some angular interval $\Delta\theta_\bq$,  and the magnitude
of the shift is below certain threshold, i.e., $q<q_{\max}$.
Obviously,   $\Delta\theta_{\mathbf{q}}$ and $q_{\max}$ also depend on $\Delta$ in a critical
manner. \cite{comment3} Approximating $\int d^{2}q$ by $\Delta\theta_{\mathbf{q}}q_{\max}^{2}$,
we resolve the $\delta$ functions and integrate over all energies to obtain

\begin{eqnarray}
\delta\sigma_{ii} & = & -\frac{e^{2}\tau_{i}^{2}T^{2}}{12}\sum_{l,m}\Delta\theta_{\mathbf{q}}|M_{\mathbf{k}_{l},\mathbf{p}_{m}}(\mathbf{q}_{\max})|^{2}\nonumber \\
\nonumber \\
 & \times & [\Delta\mathbf{v}_{i}]_{lm}^{2}\frac{k_{l}}{\mathbf{v}_{\mathbf{k}_{l}}\cdot\hat{\mathbf{k}}_{l}}\frac{p_{m}}{\mathbf{v}_{\mathbf{p}_{m}}\cdot\hat{\mathbf{p}}_{m}}\frac{1}{|\mathbf{v}_{\mathbf{k}_{l}}^{'}\cdot\hat{\mathbf{q}}|}\frac{1}{|\mathbf{v}_{\mathbf{p}_{m}}^{'}\cdot\hat{\mathbf{q}}|},\notag\\\label{delsigma3}
\end{eqnarray}
where the sum runs over all intersection points, the prime denotes
a derivative with respect to the polar angle, and $\hat{\mathbf{l}}\equiv\mathbf{l}/|\mathbf{l}|$.
The task at hand now is to find the energy dependences
of $\Delta \bv^{i}$ and $\Delta\theta_\bq$.

This task is facilitated by the geometrical construction in Fig.~\ref{fig3}(b). 
Let $\theta^{*}$ be the angle between
the normal to the FS at a point, parameterized by the angle $\theta$, and $\mathbf{q}$. As one goes around the FS contour, $\theta^{*}$ changes with $\theta$. Figure~\ref{fig3}(c) shows the dependence of $\theta^{*}$ on $\theta$ for the FS  in Fig.~\ref{fig3}(b) in the convex ($\varepsilon_F<\varepsilon_c$, dotted),  critical ($\varepsilon_F=\varepsilon_c$, dashed), and concave ($\varepsilon_F>\varepsilon_c$, solid) regimes. The dependence is monotonic for the convex FS and non-monotonic for the concave one. 
(This behavior is not specific to the particular FS considered here but is a general feature of any convex or concave contours). The oscillations are related to the rotational symmetry of the FS (six-fold
in our case; Fig. \ref{fig3}(c) shows only the domain $\theta\in[0,\pi]$). The non-monotonic
parts are centered around special (\lq\lq invariant\rq\rq\/) points passed by the $\theta^{*}(\theta)$ curves for all types of contours.  Near the invariant points, the non-monotonic part of the curve obeys a cubic equation \begin{equation}
\theta^{*}=b\theta^{3}-a(\Delta)\theta,\label{cubic}
\end{equation}
where $a(\Delta)\propto\Delta$ and $b>0$ is a constant. The energy dependences of the critical quantities can be obtained  from this equation.

To find $\Delta\theta_{\mathbf{q}}$, we note that the equation $\varepsilon_{\mathbf{k}}-\varepsilon_{\mathbf{k}-\mathbf{q}}=0$
reduces to $\mathbf{v}_{\mathbf{k}}\cdot\mathbf{q}=0$ for small $q$. This implies
that the solutions are those points on the FS where the normal to the FS is perpendicular to $\mathbf{q}$ {[}cf. Fig.~\ref{fig3}(b){]}, i.e., $\theta^{*}(\theta)=\theta_{\mathbf{q}}+\pi/2$, where $\theta_{\mathbf{q}}$ is the angle defining the direction of $\mathbf{q}$. From symmetry, if $\theta$ is a solution, so is $\theta+\pi$; therefore, one needs to consider only half the domain of $\theta$.  That only certain directions of $\mathbf{q}$ allow for more than one solution to this equation, may be appreciated by inspecting Fig.~\ref{fig3}(c), which makes it obvious that multiple roots can only occur in the regions of non-monotonicity. The interval $\Delta\theta_{\bq}$ where it happens  is then proportional to the (vertical)
width of these regions. Using Eq.~(\ref{cubic}), we find that $\Delta\theta_{\mathbf{q}}\propto\Delta\theta^{*}\propto\Delta^{3/2}$. Similarly, one can show \cite{pal} that $\Delta\mathbf{v}^{i}\propto q_{\mathrm{max}}\Delta$ and $q_{\mathrm{max}}\propto\Delta^{1/2}$ (a posteriori, this justifies the assumption of small $q$).

Substituting these results  into the expression for the conductivity, we find that $\Delta\theta_{\mathbf{q}}[\Delta\mathbf{v}_{j}]^{2}\propto\Delta^{9/2}$,
which means the prefactor of the $T^{2}$ term in the resistivity
scales as $\Delta^{9/2}$. The $T^{4}$ term is always present, as discussed
before. 
Hence, the resistivity has the following form: 
\begin{equation}
\rho=\rho_{0}+A\left(\frac{\Delta}{\varepsilon_{F}}\right)^{9/2}\Theta(\Delta)T^{2}+B\frac{T^{4}}{\varepsilon_F^2},\label{eq:result}
\end{equation}
where $\rho_{0}$ is the residual resistivity, $\theta(x)$ is the
step function, and $A$ and $B$ are material-dependent parameters (generically, $A\sim B$). A
crossover between the $T^{4}$ and $T^{2}$ regimes occurs at $T\sim\varepsilon_{F}(\Delta/\varepsilon_{F})^{9/4}\ll\varepsilon_{F}$.

 Returning to the case of  $\Delta\lesssim T$,
when both convex and concave contours are populated, it is easy to
see that the $\Delta^{9/2}$ prefactor is replaced by $T^{9/2}$,
leading to a $T^{13/2}$ term in $\rho$. This term, however, is subleading
to the $T^{4}$ one. Therefore, Eq.~(\ref{eq:result})
describes the leading $T$-dependence of the resistivity in both
situations ($|\Delta|\gg T$ and $|\Delta|\lesssim T$) near the transition. Note that the
exponents of $2$, $4$, and $9/2$ in Eq.~(\ref{eq:result}) are
universal, i.e., they are the same for an {\em arbitrary} 2D Fermi
surface with a non-quadratic energy spectrum near a convex-concave
transition.

\section{High-temperature limit}
\label{sec4}
So far, our analysis has been focused on the low-temperature limit, when the {\em ee} contribution 
to the resistivity is a correction to the {\em ei} one. From the experimental point of view, however, it is
important to understand whether the {\em ee} contribution may become larger than the {\em ei} one.
It is the case for Umklapp scattering, whose contribution grows unabated up to the temperatures comparable to the Fermi energy. The normal contribution, however, is different: it saturates in the limit when the {\em ee} relaxation time becomes
shorter then the {\em ei} one. The effect of saturation was understood already in the earlier days of the electron transport theory:\cite{herring56,keyes58} very frequent {\em ee} collision establish a quasi-equilibrium state with the drift velocity 
fixed by {\em ei} scattering. The previous analysis was, however, limited to the case when normal {\em ee} collisions
affect the resistivity via the energy dependence of the {\em ei} relaxation time.\cite{gantmakher78,levinson} In the next Section, we show that the saturation occurs even if the {\em ei} relaxation time does not depend on energy.

 \subsection{Saturation of the resistivity  in a single-band metal}
 \label{sec:sat}
We adopt the simplest model of point-like impurities with energy-independent scattering time, when
the BE is given by Eq.~(\ref{tra}).
The {\em ee} collision integral in (\ref{tra}) can be viewed as a linear operator $\hat{I}_{\mathrm{ee}}$ acting on the non-equilibrium part of the distribution function $f^{(1)}_{\bk}\equiv f_\bk-n_\bk$
\begin{eqnarray}\label{collision operator}
I_{\mathrm{ee}}[f^{(1)}]({\bk}) \equiv \sum_{\bk'}I_{\mathrm{ee}}(\bk,\bk')f^{(1)}_{\bk'},
\end{eqnarray}
The non-Hermitian matrix operator $\hat{I}_{\mathrm{ee}}$ can be represented
in terms of its left, $\tilde{\Phi}^{\lambda}$, and right, $\Phi^{\lambda}$,
eigenstates as
\begin{eqnarray}\label{M-expansion}
\hat{I}_{\mathrm{ee}} = \frac{1}{\tau_{\mathrm{ee}}^*}\sum_{\lambda}|\Phi^{\lambda}\rangle \lambda \langle \tilde{\Phi}^{\lambda}|\, \, ,
\end{eqnarray}
where  $\tau_{\mathrm{ee}}^*$ is the effective {\em ee} scattering time. The right and left states constitute an orthonormal basis:
\begin{eqnarray}\label{orthogonality}
\langle \tilde{\Phi}^{\lambda'}|\Phi^{\lambda}\rangle \equiv \frac{1}{\mathcal{V}}\sum_{\bk}
\tilde{\Phi}^{\lambda'}(\bk)\, \Phi^{\lambda}(\bk) = \delta_{\lambda, \, \lambda'} \,,
\end{eqnarray}
where $\mathcal {V}$ is the system volume in the D-dimensional space. A general solution of Eq.~(\ref{tra}) can be expanded over the complete basis as 
\begin{eqnarray}\label{superposition}
f^{(1)}_\bk = \sum_{\lambda}c_{\lambda} \Phi^{\lambda}(\bk)\, ;
\end{eqnarray}
substituting this form into Eq.~(\ref{tra}), we obtain an equation for the coefficients $c_{\lambda}$:
\begin{eqnarray}\label{c}
\left[\frac{1}{\ti} + \frac{\lambda}{\tau_{\mathrm{ee}}^*}\right]c_{\lambda} = e\left\langle \tilde{\Phi}^{\lambda}\left| \right. \bv_\bk\cdot {\bf E} n'_\bk\right\rangle \, .
\end{eqnarray}
Only the zero mode ($\lambda =0$) contribution survives in the limit of
$1/\tau_{\mathrm{ee}}^* \rightarrow \infty$, so the solution in the high-$T$ regime is given by
\begin{eqnarray}\label{solution}
\left.f^{(1)}_\bk\right\vert_{T\to\infty}\!\! = e\ti \Phi^{\lambda=0}(\bk)\frac{1}{\mathcal{V}}\sum_{\bk'}\tilde{\Phi}^{\lambda=0}(\bk')
\bv_{\bk'}\cdot{\bf E}n'_{\bk}.
\label{4.6}
\end{eqnarray}
It is not difficult to see that the right and left \emph{zero} modes of $\hat{I}_{\mathrm{ee}}$ are
\begin{eqnarray}\label{zero-modes}
|\Phi^{\lambda=0}_{i}(\bk)\rangle = - C_{i} k_{i} n'_\bk\,\,\, \,\, ; \,\,\, \,\, \langle \tilde{\Phi}^{\lambda=0}_{i}(\bk)| = \tilde{C}_{i} k_{i}.
\end{eqnarray}
Indeed,
\beq
\langle\tilde{\Phi}^{\lambda=0}\vert
\hat{I}_{\mathrm{ee}}
\propto \sum_{\bk}\bk\,I_{\mathrm{ee}}(\bk,\bk') = 0
\eeq
due to momentum conservation; while
\beq
\hat{I}_{\mathrm{ee}}| \Phi^{\lambda=0}\rangle \propto \sum_{\bk'}I_{\mathrm{ee}}(\bk,\bk')
\bk'n'_{\bk'} = 0\eeq
because the collision integral, evaluated for equilibrium distribution functions, remains to be equal to zero
if all energies are shifted as $\ek \rightarrow \ek + {\bf u}\cdot\bk$ with ${\bf u}$ being an arbitrary $\bk$-independent vector.
The zero modes 
form a $D$-dimensional subspace labeled by the Cartesian indices
$i = 1, 2, \dots D$ [
summation 
over these indices
is implied in (\ref{superposition}) and (\ref{4.6})].
The scalar product (\ref{orthogonality})  of the zero-modes (\ref{zero-modes})
is
\begin{eqnarray}\label{norm}
\langle \tilde{\Phi}^{\lambda=0}_{i}|\Phi^{\lambda=0}_{j}\rangle& =&
\frac{\tilde{C}_{i}C_{j}}{\mathcal{V}}\sum_{\mathbf{k}}k_{i}
k_{j}(-n'_{\mathbf{k}}) \notag\\
&&\equiv \tilde{C}_{i}C_{j}\nu(E_F)\langle k_{i}k_{j}\rangle
  \, ,
\end{eqnarray}
where 
\begin{eqnarray}\label{averaging}
\langle F
\rangle \equiv \frac{1}{\nu(\varepsilon_F)\mathcal{V}}
\sum_{\bk}F(\bk)(-n'_\bk)
  \, .
\end{eqnarray}
The scalar product (\ref{norm})
is diagonal
in the coordinate system associated with the principal axes
of the quadratic form
$\langle k_{i}k_{j}\rangle$; 
normalization is ensured by
choosing $\tilde{C}_{i}C_{i} =
[\nu(E_F)\langle k^2_{i}\rangle]^{-1}$.
Using these properties,  we reduce Eq.~(\ref{solution}) to
\begin{eqnarray}\label{strong solution}
f^{(1)}_{\bk} &=& e\ti\sum_{i}\tilde{C}_{i}C_{i}k_{i}(-n'_\bk) \frac{1}{\mathcal{V}}
\sum_{\bk'}k'_{i} \bv_{\bk'}\cdot{\bf E}n_{\bk'}' \notag\\
&&=
-e\ti\sum_{i, j} k_{i}(-n'_\bk) \frac{\langle k_{i}v_{j}\rangle}{\langle k^2_{i}\rangle}E_{j}
 \, .
\end{eqnarray}
Finally, we obtain the conductivity tensor in the high-$T$ limit as
\begin{eqnarray}\label{conductivity-strong}
\left.\sigma
_{ij}\right|_{T\to\infty} = 2e^2\nu(E_F)\ti\sum_{\l}
\langle v_{i}k_{\l}\rangle
\frac{\langle k_{\l}v_{j}\rangle}{\langle k^2_{\l}\rangle}
\, .
\label{highT}
\end{eqnarray}
In the opposite limit of low temperatures, the standard expression reads
\begin{eqnarray}\label{conductivity-strong}
\left.\sigma_
{ij} \right\vert_{T\to 0}= 2e^2\nu(E_F)\ti\langle v_{i}v_{j}\rangle
\, .
\end{eqnarray}

In contrast to the case of energy-dependent $\ti$, when the low- and high-temperature limits of the conductivity
differ in how $\ti$ is averaged over the energy,~\cite{gantmakher78,levinson} these limits in our case differ in how the conductivity is averaged over the FS.
Naturally, the two limits coincide for the Galilean-invariant case. 
Notice that saturation holds for any dimensionality and shape of the FS, i.e., regardless of whether the temperature dependences of the resistivity starts with a $T^2$ or $T^4$ term at low temperatures, it will saturate at high temperatures.
In reality, of course, other scattering mechanisms, such as electron-phonon scattering, will mask the resistivity saturation.

 If the FS is not abnormally anisotropic, the low- and high-$T$ limits are of the same order, which means that a true $T^2$-scaling regime does not have room to develop. A mechanism in which such a regime is possible is considered in the next Section.

\subsection{Two-band model: Scaling regime}
\label{sec:2band}
In this Section, we consider
a simple model of a two-band metal with impurities. 
Since normal {\em ee} collisions affect the resistivity for any multiply-connected FS, we consider the simplest
 case of two bands with quadratic
dispersions, $\varepsilon_{\mathbf{k}}^{\left(1,2\right)}=k^{2}/2m_{1,2}$,
and, in general, different impurity scattering times, $\tis$ and
$\tid.$  We consider only the inter-band interaction  (the intra-band one
drops out in this case anyway) and neglect processes in which electrons are transferred from one band to another. 
The BE for this model in 2D can be solved exactly by generalizing the method of  Appel and Overhauser \cite{appel78}(see Appendix \ref{app:twoband}) with the result
\beq
\rho(T)=\frac{\pi}{e^2\varepsilon_F}\frac{\frac{1}{\tau_{\mathrm{i}1}\tau_{\mathrm{i}2}}
+\frac{1}{\tau_{\mathrm{ee}}(T)}\left(\frac{1}{\tau_{\mathrm{i}1}}\frac{m_1}{m_2}+\frac{1}{\tau_{\mathrm{i}2}}\frac{m_2}{m_1}\right) }{\frac{1}{\tau_{\mathrm{i}1}} +\frac{1}{\tau_{\mathrm{i}2}}+\frac{1}{\tau_{\mathrm{ee}}(T)}\left(2+\frac{m_1}{m_2}+\frac{m_2}{m_1}\right)},
\label{2band}\eeq
where
\begin{widetext}
\bea
\frac{1}{\tau_{\mathrm{ee}}(T)}
=
\frac{\sqrt{m_{1}m_{2}}}{2T\varepsilon_{F}^{2}}\int^{\infty}_{-\infty}d\omega\int^{2k_F^{\min}}_0\frac{dqq}{2\pi}
\frac{W\left(\bq,\omega\right)}{\sqrt{1-(q/k_{F1})^2}\sqrt{1-(q/2k_{F2})^2}}
\omega^2N\left(\omega\right)\left[N\left(\omega\right)+1\right]
\label{tau_gen}
\eea
\end{widetext}
and 
$k_F^{\min}\equiv\min\{k_{F1},k_{F2}\}$.
The result for the resistivity follows already from the equations of motion\cite{levinson}
\begin{eqnarray}
m_1\frac{d\bv_{1}}{dt} & = &- e\mathbf{E}-\frac{m_{1}\mathbf{v}_{1}}{\tis}-
\eta n_{2}
(\mathbf{v}_{1}-\mathbf{v}_{2})=0\notag\\
m_2\frac{d\bv_{1}}{dt}& = & -e\mathbf{E}-\frac{m_{2}\mathbf{v}_{2}}{\tid}-\eta n_{1}(\mathbf{v}_{1}-\mathbf{v}_{2})=0,
\end{eqnarray}
if the phenomenological \lq\lq friction coefficient\rq\rq\/ $\eta$ is expressed via the microscopic scattering time as
$\eta=\pi/\varepsilon_{F}\tau_{\mathrm{ee}}(T)$.
 
An interesting case is when the masses are significantly different (as would be the case for a metal with partially occupied {\em s} and {\em d} bands~\cite{ziman}), e.g., $m_2\gg m_1$ and consequently, $\tid\ll\tis$.
At $T\to 0$, the two bands conduct in parallel, and the total resistivity is dominated by that of the lighter band
\beq
\rho(0)=\frac{\pi }{e^2\varepsilon_F}\frac{1}{\tis+\tid}\approx \frac{\pi}{e^2\varepsilon_F\tis}.
\eeq
At $T\rightarrow\infty,$
the resistivity saturates at a value determined by the resistivity
of the heavy band 
\begin{equation}
\rho \left( \infty \right) =\frac{\pi }{e^{2}\varepsilon_F}\left(\frac{1}{\tis}\left(\frac{m_1}{m_2}\right)^2+\frac{1}{\tid}\right)\approx \frac{\pi }{e^{2}\varepsilon _{F}\tid}.
\end{equation}
Therefore, the $T=0$ and $T=\infty$ limits now differ significantly 
\beq
\frac{\rho\left(\infty\right)}{\rho\left(0\right)}=\frac{\tis}{\tid}=\frac{m_{2}}{m_{1}}\gg1.
\eeq
The scaling regime, in which 
\beq
\rho\left(T\right)\approx\frac{\pi}{e^{2}}\frac{m_{2}}{m_{1}}\frac{1}{\varepsilon_{F}\tau_{\mathrm{ee}}(T)}
\eeq
occurs in a wide temperature interval  $T_{l}$ $\ll T\ll$ $T_{h}$, the boundaries of which are defined by 
\begin{eqnarray}
\frac{1}{\tau_{\mathrm{ee}}\left(T_{l}\right)} = \frac{1}{\tis}\frac{m_{1}}{m_{2}};\;\frac{1}{\tau_{\mathrm{ee}}\left(T_{h}\right)} & = & \frac{1}{\tid}\frac{m_{1}}{m_{2}
}.
\end{eqnarray}

This model can also be applied to the QPT, in which case it is natural to assume that critical fluctuations occur only in the heavy band. Consequently, the effective interaction is obtained from Eq.~(\ref{hm1}) by replacing $v_F\to v_{F2}$ and $\nu_F\to m_2/2\pi$. Computing the integral (\ref{tau_gen}) for $\delta=0$ and $q\ll k_F^{\min}$, we find the effective scattering rate in the
NFL regime 
\begin{equation}
\frac{1}{\tau_{\mathrm{ee}}(T)}=\frac{16\pi^{3}\zeta(4/3)}{81\Gamma(2/3)}\frac{\sqrt{m_{1}m_{2}}}{\varepsilon_{F}^{2}}\frac{1}{m^2_2a^{2}}\left(\frac{v_{F2}}{a}\right)^{2/3}T^{4/3},\label{nfl}
\end{equation}
where $\zeta(x)$ and $\Gamma(x)$ are the Riemann and $\Gamma$-functions, correspondingly.
In this scenario, normal {\em ee} collisions do lead to a real scaling regime in the resistivity with an exponent
given by \lq\lq naive\rq\rq\/ power-counting argument.
\section{\label{sec5} Limitations of the Boltzmann-equation approach}
The semiclassical BE does not capture two types of effects. The first type--quantum-- results from quantum interference
between {\em ee} and {\em ei} scattering; the second one --classical--from correlations in the electron flow patterns produced by different impurities. In this section, we discuss the limits of validity of the semiclassical approach focusing on the 2D case.

\subsection{Quantum-interference effects}
\label{sec:qint}
\subsubsection{Fermi-liquid regime}
\begin{figure}
\includegraphics[width=0.45\textwidth]{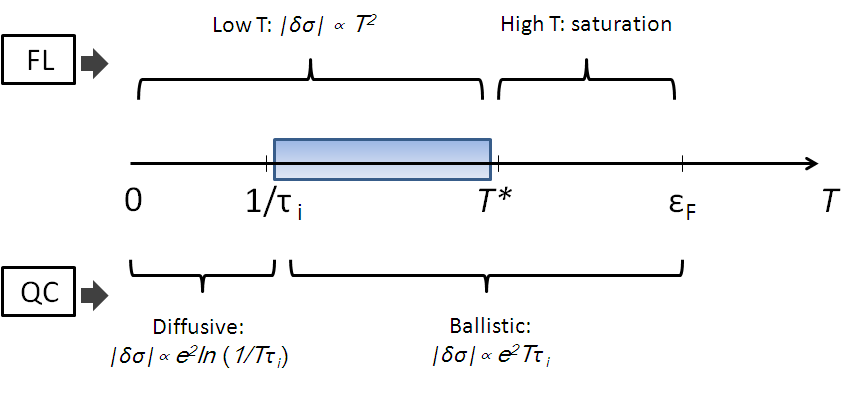} \caption{Different temperatue regimes for Fermi-liquid (FL)  and quantum-interference (QC) corrections to the conductivity. The shaded region on the temperature scale is the regime where the FL ($T^2$) correction is dominant.}
\label{fig:temperatures}
\end{figure}

Recall that the FL-like contribution to the resistivity, discussed in this paper, behaves as $T^2$ (or $T^4$, if there is approximate integrability) in the low-temperature regime, defined by $1/\tau_{\mathrm{ee}}\ll 1/\ti$ and saturates in the high-temperature regime, defined by  $1/\tau_{\mathrm{ee}}\gg 1/\ti$.  In non-integrable systems, $1/\tau_{\mathrm{ee}}=gT^2/\varepsilon_F$,  where $g$ is the dimensionless coupling constant. In a generic FL,  $g\sim 1$ and the crossover between the two limits occurs at $T^{\star}=\sqrt{\varepsilon_F/\ti}$. For a good metal, $\varepsilon_F\ti\gg 1$ so that $1/\tau_i\ll T^{\star}\ll \varepsilon_F$. In case of quantum corrections (QC), the scale that differentiates between low and high temperatures, i.e., between the diffusive and ballistic regimes, is $T_{\mathrm{DB}}=1/\ti$. For $T\ll T_{\mathrm{DB}}$, one is in the diffusive limit, characterized by a logarithmically divergent Altshuler-Aronov correction;~\cite{altshuler:85} with all coupling constants being of order one,  $|\delta\sigma|/\sigma_D\sim \ln(1/T\ti)/\varepsilon_F\ti$, where $\sigma_D=e^2\varepsilon_F\ti/\pi$ is the Drude conductivity. For $T\gg T_{\mathrm{DB}}$, one is in the ballistic limit, where the correction scales linearly with $T$: $|\delta\sigma|/\sigma_D\sim T/\varepsilon_F$ (Ref.~\onlinecite{zala:01}). Apart from the interaction correction, there is a also a weak-localization correction $-\delta\sigma_{\mathrm{WL}}/\sigma_D\sim \ln(\tau_{\phi}/\ti)/\varepsilon_F\ti$, where $\tau_{\phi}$ is the phase-breaking time, a precise form of which depends on whether one is in the diffusive or ballistic limits: in the former, $1/\tau_{\phi}\sim T\ln( \varepsilon_F\ti)/ \varepsilon_F\ti$; in the latter, $1/\tau_{\phi}\sim 1/\tau_{\mathrm{ee}}$. In the diffusive limit, the weak-localization correction is similar to the Altshuler-Aronov result, differing only in the prefactor. In the ballistic limit, the weak localization correction is smaller than the interaction correction by a factor of $\ln(T^*/T)/T\ti$. Therefore, the correct order of magnitude for the quantum-interference correction is still given by the interaction correction both in the diffusive and ballistic limits.
Comparing the FL-contribution $-\delta\sigma_{\mathrm{FL}}/\sigma_{D}\sim T^2\tau_i/\varepsilon_F$ to the quantum corrections, we find that $|\delta\sigma_{\mathrm{QC}}/\delta\sigma_{\mathrm{FL}}|\sim \mathrm{ln}(1/T\tau_i)/T^2\tau_i^2\gg 1$ and $|\delta\sigma_{\mathrm{FL}}/\delta\sigma_{\mathrm{QC}}|\sim T\tau_i\gg 1$ in the diffusive and ballistic limits, correspondingly.
Therefore, it is meaningful to consider the FL contribution and neglect quantum-interfence processes in the ballistic but not in the diffusive limit. The interplay of different mechanisms is shown schematically in Fig.~\ref{fig:temperatures}.

In the integrable case, the $T^2$ term in the resistivity vanishes and the FL correction scales as $|\delta\sigma_{\mathrm{FL}}|/\sigma_D\sim
T^4\ti/\varepsilon_F^3$. In this case, the FL correction dominates over the quantum one only at temperatures well above  the diffusion-ballistic crossover: $T\gg (\varepsilon_F\ti)^{2/3}T_{\mathrm{DB}}\gg T_{\mathrm{DB}}$. It is worth noting that the discussion of the experimental observations of quantum corrections in the ballistic regime has been so far limited to 2D electron gases in Si and GaAs heterostructures \cite{kravchenko}, with essentially circular FSs and almost parabolic spectra, where the FL contribution is expected to be very small. The FL contribution, however, is expected to play a dominant role in 2D systems with highly anisotropic FSs and non-parabolic spectra, such as the surface state in the topological insulators of the Bi$_2$Te$_3$ family, discussed in Sec.~\ref{sec:bi2te3}.

\subsubsection{Non-Fermi--liquid regime}
In this section, we describe the interplay between the quantum-interference and direct {\em ee} contributions to the resistivity in the NFL--regime of a ferromagnetic quantum phase.  For simplicity, we assume that the integrability is broken already in a single-band case by sufficiently strong concavity of the FS. In this case, one can simply calculate the transport time for scattering at critical spin fluctuations, described by the propagator (\ref{hm1}) with $\delta=0$, and substitute the result into the Drude formula. \cite{schofield:99} This gives
$-\delta\sigma_{\mathrm{NFL}}/\sigma_D=T^{4/3}\ti/T_0^{1/3}$, where $T_0\sim (v_F/a)(k_Fa)^9$ (we remind that $k_Fa\gg 1$ is a control parameter of the HMM model). The temperature above which the NFL contribution saturates is now given by
$T^*\sim \varepsilon_F^{1/4}(k_Fa)^2/\ti^{3/4}$. The main difference between the FL-- and NFL--regimes is that quantum criticality changes 
space-time (or energy-momentum) scaling: while $\omega\propto q$ in a FL, $\omega\propto q^Z$ near QCP, where $Z$ is the dynamical exponent ($Z=3$ in the HMM model). Therefore, the temperature of the diffusive-ballistic crossover, determined by the condition ${\bar q}v_F\ti\sim 1$, where ${\bar q}\sim (T/v_Fa^2)^{1/3}$ is a typical value of the momentum transfer in an {\em ee} collision, is replaced by $T_{\mathrm{DB}}\sim (a/v_F\ti)^2/\ti$ (Ref.~\onlinecite{paul:05}).  In a clean system, where $a/v_F\ti\ll 1$, $T_{\mathrm{DB}}$ is significantly smaller than in a FL, where $T_{\mathrm{DB}}\sim 1/\ti$, so that the ballistic regime continues down to much lower temperatures compared to the FL case, and we limit our analysis to this regime. (Lowering of $T_{\mathrm{DB}}$ in the vicinity of a ferromagnetic QCP becomes noticeable already in the FL regime. \cite{zala:01}) Another consequence of quantum-critical scaling is that the quantum correction in the ballistic regime
near a QCP behaves as $T^{1/3}$ as opposed to $T$:  $-\delta\sigma_{\mathrm{QC}}/\sigma_D\sim \left(T/\varepsilon_F\right)^{1/3}(ak_F)^{4/3}$ (Ref.~\onlinecite{paul:05}). Comparing the NFL and QC contributions, we find that the NFL contribution dominates only at  $T\gg {\tilde T}$, where  ${\tilde T}\sim(ak_F)^4/\ti\sim T_{\mathrm{DB}}(\varepsilon_F\ti)^2(k_Fa)^2\gg T_{\mathrm{DB}}$. A relative weakness of 
of the NFL contribution is due to small-angle scattering at long-wavelength critical fluctuations.

\subsection{Viscous contribution to the resistivity}
\label{sec:visc}
The statement that {\em ee} interaction does not contribute to the resistivity
of a Galilean-invariant FL (cf. Sec.~\ref{sec:general} )
seems to contradict an intuitive notion that it is the viscosity of a liquid
that defines its rate of flow. In a certain regime, indeed, the resistivity does depend on the viscosity of the electron liquid.~\cite{gurzhi64,hruska02}
This effect is not taken into account by the standard BE which neglects not only quantum but also classical correlations between scattering events. The "viscous" contribution occurs at high enough temperatures, when  the mean free path due to the {\em ee} interaction, $l_{\mathrm{ee}}=v_F\tau_{\mathrm{ee}}$, is smaller than at least the average distance between impurities, $1/N^{1/D}_i$, where $N_i$ is the number density of impurities.~\cite{hruska02} In this regime, the {\em ei} mean free path $l_{\mathrm{i}}\gg 1/N^{1/D}_i\gg l_{\mathrm{ee}}$ is the largest scale of the problem, which implies that the FL contribution--even if allowed due to anisotropy of the FS--has already saturated of at a value comparable to the Drude resistvity (cf. Sec.~\ref{sec4}). Barring phonons, the viscous contribution is the only source of the $T$ dependence in this regime.

To estimate the magnitude of the viscous contribution, we consider, following Ref.~\onlinecite{hruska02}, a flow of the electron liquid through a random array of spherical
impurities. First, the impurity radius $R$ is assumed to be much larger than $l_{\mathrm{ee}}$,
so that  a hydrodynamic description
is applicable at all lengthscales. The force on one electron
from all impurities is just the Stokes
force $F_{S}\sim (N_{i}/N)\mu uR$, where $\mu $ is the dynamic viscosity, $u$
is the flow velocity, and $N$ is the electron number density.
[An exact value of the numerical coefficient in $F_S$ depends on the boundary conditions for the velocity at the surface of the sphere\cite{fluid_dynamics}  but will not be needed here.]
In steady state,
$F_{S}=eE$, which yields $u\sim eEN/N_{i}\mu R$, and thus the viscous contribution to the resistivity is given by
$\delta\rho_{v}\sim N_{i}\mu R/e^{2}N^{2}$.
In a
FL, $\delta\rho_v\propto \mu\sim mNv_Fl_{\mathrm{ee}}\propto1/T^{2}$; thus the viscous correction is of the {\em insulating} sign.
On the other hand, the Drude resistivity resulting from scattering off the same impurities is
$\rho_{D}=m/e^2N\ti\sim mv_{F}N_{i}R^{2}/e^{2}N$.
The viscous contribution is smaller than the Drude resistivity within the hydrodynamic regime:
$\delta\rho_{v}/\rho_{D}\sim l_{ee}/R\ll 1$.

In 2D, the Stokes force from  a disk-like impurity depends on $R$ only logarithmically:
$F_{S}=4\pi\mu u/L$,
where $L=\ln(3.70\mu/RmNu)$.~\cite{fluid_dynamics} However, $1/\ti\sim v_{F}N_{i}R$
also contains $R$ instead of $R^2$, so that the ratio $\delta\rho_v/\rho_D$
is the same (up to a logarithm) as in 3D.

The situation is somewhat different for small impurities, because a force exerted
by a small ($R\ll l_{ee}$) sphere on a rarified gas depends not on
the viscosity but on the gas-solid accommodation coefficients. \cite{phys_kin} In 2D, the situation is further
complicated vy the Stokes paradox.~\cite{fluid_dynamics}  Hruska and Spivak~\cite{hruska02}
showed that the viscous correction for small impurities in 2D is given by
 \begin{equation}
-\frac{\delta\rho_{v}}{\rho_D}\sim\frac{a}{l_{ee}}\ln\left(\frac{1}{N_{i}^{1/2}l_{ee}}\right),
\end{equation}
where $a$ is the impurity scattering length and $a\ll l_{\mathrm{ee}}\ll 1/N_i^{1/2}$ . Because of a large logarithmic factor, $\delta\rho_{v}$ can, in principle, be comparable to $\rho_D$.

\section{\label{sec6} Conclusions}

The main purpose of this paper was to analyze the
effect of \emph{ee} interactions on the resistivity in the situation when
Umklapp scattering of electrons can be neglected. Such a situation arises, e.g., in low-carrier density
materials, as well as in metals near a $q=0$
QPT, where the effective interaction is of a long range. In such cases,
the conventional $T^{2}$  dependence (or its $T^{\frac{D+2}{3}}$ analog in the NFL region near a QPT)
 of the resistivity on temperature
is not guaranteed. Whether it
is present  depends on 1) dimensionality, 2) shape, and 3)
topology of the FS. If the FS is quadratic or isotropic,
there is no $T^{2}$ contribution to
the resistivity. However, anisotropy is not sufficient to guarantee
the $T^{2}$ dependence. In the case of a convex and simply connected FS in 2D, there is no $T^{2}$ dependence
either. In such cases, the leading temperature dependence on resistivity
due to \emph{ee} interactions is $T^{4}$ in the FL region and $T^{(D+8)/3}$ in the NFL region. Also, if the FS changes
its shape from convex to concave as a function of the filling fraction, the resistivity follows a
universal scaling form near the convex-concave transition. In all other cases, the $T^{2}$ (or $T^{\frac{D+2}{3}}$)
behavior is allowed, albeit only as a correction to the Drude resistivity.
However, a true scaling regime (when the {\em ee} contribution is larger then the {\em ei} one) is possible
for a quantum-critical two-band metal with substantially different band masses. 
Since a quantum-critical behavior is observed typically in multi-band metals with partially occupied {\em d} bands, we conjecture that the $5/3$ scaling of the resistivity observed in 3D ferromagnets \cite{exp_fm} and subquadratic scaling in a quasi-2D metamagnet Sr$_3$Ru$_2$O$_7$ \cite{metamagn} is due to the interaction between light and heavy carriers in these materials.

\acknowledgements
Regrettably, this paper could not been discussed with Yehoshua B. Levinson (1932-2008), who contributed immensely to the field of electron transport in solids. Nevertheless, one of the authors (D.L.M.) had a privilege to learn firsthand about many of the issues addressed here from Levinson more than two decades ago. The reader may also notice that this paper frequently cites the classic monograph by Gantmakher and Levinson (Ref.~\onlinecite{levinson}), which in and of itself, speaks about Levinson's legacy.

We would like to thank A. Chubukov, who collaborated with us on many aspects of this work. Stimulating discussions with A. Finkelstein, A. Kamenev, M. Kargarian, S.-S. Lee, D. Loss, I. Paul, C. P{\'e}pin, M. Reizer, S. Sachdev, B. Spivak, and A. Varlamov are gratefully acknowledged . This work was supported by RFBR-12-02-00100 (V.I.Y.) and NSF-DMR-0908029 (D.L.M.)
\appendix
\section{\label{app:twoband}{Two-band model in 2D}}
\begin{widetext}
In this Appendix we derive Eq. ~(\ref{2band}).  The two coupled BEs (1 and 2
refers to the bands 1 and 2) read
\begin{eqnarray}
-e\mathbf{v}_{1}\cdot {\boldsymbol \nabla}_{\bk_1}f_{1} & = & -I_{\mathrm{ee}}^{12}\left[f_{1},f_{2}\right]-\frac{f_{1}\left(\mathbf{k}_{1}\right)-n_{1}}{\tis}\notag\\
-e\mathbf{v}_{2} \cdot {\boldsymbol \nabla}_{\bk_1}f_{2}& = & -I_{\mathrm{ee}}^{21}\left[f_{1},f_{2}\right]-\frac{f_{2}\left(\mathbf{k}_{2}\right)-n_{2}}{\tid}
\end{eqnarray}
where $I_{\mathrm{ee}}^{ij}$ is the electron-electron collision integral for
scattering between two electrons from the $i$th and $j$th band, and $n_{1,2}$ are the equilibrium distributions.
Since in our model each of the bands is Galilean-invariant on its own, the intra-band {\em ee} interaction cannot affect the resistivity,
and the corresponding parts of the collision integrals are not written down.
Linearizing $f_{\alpha}$ ($\alpha=1,2$) in the same way as for the single-band case 
\beq
f_{\alpha}=n_{\alpha}+n_{\alpha}\left(1-n_{\alpha}\right)g_{\alpha}=n_{\alpha}-Tn_{\alpha}^{\prime}g_{\alpha},
\eeq
we obtain 
\bea
-e\mathbf{v}_{1}\mathbf{\cdot E}n_{1}^{\prime} & = & -I_{\mathrm{ee}}^{12}\left[g_{1},g_{2}\right]+\frac{Tn_{1}^{\prime}g_{1}}{\tis},\notag\\
-e\mathbf{v}_{2}\mathbf{\cdot E}n_{2}^{\prime} & = & -I_{\mathrm{ee}}^{21}\left[g_{1},g_{2}\right]+\frac{Tn_{2}^{\prime}g_{2}}{\tid}
\eea
where 
\begin{eqnarray}
I_{\mathrm{ee}}^{12} & = & \int\frac{d^{2}k_{2}}{\left(2\pi\right)^{2}}\int\frac{d^{2}k_{1}^{\prime}}{\left(2\pi\right)^{2}}\int\frac{d^{2}k_{2}^{\prime}}{\left(2\pi\right)^{2}}W\left(\bk-\bkp,\varepsilon_{\bk_1}-\varepsilon_{\bk_1'}\right)n_{1}\left(\mathbf{k}_{1}\right)n_{\mathbf{2}}\left(\mathbf{k}_{2}\right)\left(1-n_{\mathbf{1}}\left(\mathbf{k}_{1}^{\prime}\right)\right)\left(1-n_{\mathbf{2}}\left(\mathbf{k}_{2}^{\prime}\right)\right)\notag\\
 &  & \times\left[g_{1}\left(\mathbf{k}_{1}\right)+g_{2}\left(\mathbf{k}_{2}\right)-g_{1}\left(\mathbf{k}_{1}^{\prime}\right)-g_{2}\left(\mathbf{k}_{2}^{\prime}\right)\right]\delta\left(\varepsilon_{\mathbf{k}_{1}}+\varepsilon_{\mathbf{k}_{2}}-\varepsilon_{\mathbf{k}_{1}^{\prime}}-\varepsilon_{\mathbf{k}_{2}^{\prime}}\right)\delta\left(\mathbf{k}_{1}\mathbf{+k}_{2}\mathbf{-k}_{1}^{\prime}-\mathbf{k}_{2}^{\prime}\right)
\end{eqnarray}
and $I_{\mathrm{ee}}^{21}$ differs from $I_{\mathrm{ee}}^{21}$ in that the first integral goes over $\bk_1$ instead of $\bk_2$.
We seek for a solution in the following form 
\beq
g_{\alpha}=ec_{\alpha}\left(\mathbf{v}_{\alpha}\cdot\mathbf{E}\right)/T,
\eeq
where $c_{\alpha}$ are the constants that are to be determined. Multiplying
the first and second BEs by $\mathbf{v}_1\cdot{\bf E}$ and $\mathbf{v}_{2}\cdot{\bf E}$
and integrating over $\mathbf{k}_{1}$ and $\mathbf{k}_{2}$, correspondingly, we obtain for the left-hand sides  
\bea
-e\int\frac{d^{2}k_{\alpha}}{\left(2\pi\right)^{2}}\left(\mathbf{v}_{\alpha}\cdot {\bf E}\right)^{2}n_{\alpha}^{\prime}=\frac{m_{\alpha}}{4\pi}v_{F\alpha}^{2}eE^2
\eea
Similarly, the {\em ei} collision integrals, integrated over the corresponding momenta, reduce to
\beq
\int\frac{d^{2}k_{\alpha}}{\left(2\pi\right)^{2}}\frac{Tn_{\alpha}^{\prime}g_{\alpha}}{\tau_{\mathrm{i}\alpha}}\bv_{\alpha}\cdot{\bf E}=-c_{\alpha}\frac{m_{\alpha}}{4\pi}\frac{v_{F\alpha}^{2}}{\tau_{i\alpha}}eE^2
\eeq
For a FL, the scattering probability may be assumed to depend
only on the momentum transfer but not energy transfer. In this case, the integrals of {\em ee} collision integrals multiplied over $\bv_{\alpha}\cdot {\bf E}$ reduce to 
\begin{eqnarray}
 \int\frac{d^2k_1}{(2\pi)^2}\bv_1\cdot {\bf E}I_{ee}^{12} = eE^2R\frac{m_2}{v_{F1}v_{F2}}
  \left(\frac{c_{1}}{m_{1}}-\frac{c_{2}}{m_{2}}\right),
 \label{int_iee}
\end{eqnarray}
where 
\bea
R\equiv  \frac{\pi}{3}T^{2}
\int^{2k_F^{\min}}_0 dqq\frac{W\left(q\right)}{\sqrt{1-\left(\frac{q}{2k_{F1}}\right)^2}\sqrt{1-\left(\frac{q}{2k_{F2}}\right)^2}}
\eea
with $k_F^{\min}\equiv\min\{k_{F1},k_{F2}\}$, and $\int_{\bk_{2}}\bv_2\cdot{\bf E}I_{ee}^{21}$ differs from (\ref{int_iee}) by a factor of $m_1/m_2$. [Integration over energies was performed with the help of Eq.~(\ref{ident}).]
Solving the system of linear equations $c_{1}$ and $c_{2}$
\begin{eqnarray}
\frac{m_{1}}{4\pi}v_{F1}^{2} & = & -
\frac{c_{1}}{\tis}\frac{m_{1}}{4\pi}v_{F1}^{2}-
R\frac{m_{2}}{v_{F1}v_{F2}}\left[\frac{c_{1}}{m_{1}}-\frac{c_{2}}{m_{2}}\right]
\notag\\
\frac{m_{2}}{4\pi}v_{F2}^{2} & = & -\frac{c_{2}}{\tid}\frac{m_{2}}{4\pi}v_{F2}^{2}+
R\frac{m_{1}}{v_{F1}v_{F2}}\left[\frac{c_{1}}{m_{1}}-\frac{c_{2}}{m_{2}}\right],
\end{eqnarray}
we find 
\begin{eqnarray}
c_{1} & = & -\frac{1/\tid+\left(1/\tau_{\mathrm{ee}}\right)\left(m_{1}/m_{2}+1\right)}{\frac{1}{\tis\tid}+\frac{1}{\tau_{\mathrm{ee}}}\left(\frac{1}{\tis}\frac{m_{1}}{m_{2}}+\frac{1}{\tid}\frac{m_{2}}{m_{1}}\right)}\notag\\
c_{2} & = & -\frac{1/\tis+\left(1/\tau_{\mathrm{ee}}\right)\left(m_{2}/m_{1}+1\right)}{\frac{1}{\tis\tid}+\frac{1}{\tau_{\mathrm{ee}}}\left(\frac{1}{\tis}\frac{m_{1}}{m_{2}}+\frac{1}{\tid}\frac{m_{2}}{m_{1}}\right)},
\end{eqnarray}
where the effective {\em ee} scattering time was introduced as  
\begin{eqnarray}
\frac{1}{\tau_{\mathrm{ee}}} & \equiv & 4\pi R\frac{1}{\sqrt{m_{1}m_{2}}}\frac{1}{v_{F1}^{2}v_{F2}^{2}}.
\label{tau}
\end{eqnarray}
Using that the Fermi energy is same for both bands, i.e., that $
v_{F1}/v_{F2}=\sqrt{m_2/m_1}$, we cast Eq.~(\ref{tau}) into a different form
\bea
\frac{1}{\tau_{\mathrm{ee}}(T)}=
\frac{\pi^2}{3}T^2
\frac{\sqrt{m_{1}m_{2}}}{\varepsilon_{F}^{2}}\int^{2k_F^{\min}}_0 dqq
\frac{W\left(\bq,0\right)}{\sqrt{1-(q/2k_{F1})^2}\sqrt{1-(q/2k_{F2})^2}}.
\label{tau_FL}
\eea
Once $c_{1,2}$ are found, one readily finds the electric current and arrives at the expression for
the resistivity quoted in Eq.~(\ref{2band}).  An explicit expression for $1/\tau_{\mathrm{ee}}$ was derived in Ref.~\onlinecite{murzin98} for
a special case of the \lq\lq\ overscreened\rq\rq\/ Coulomb potential. Our result for this case coincides with that in Ref.~\onlinecite{murzin98} up to a numerical coefficient.

In general, the {\em ee} scattering probability depends not only on $q$ but also on $\omega$. In this case, the integrals over energies cannot be performed in a general form, and the expression for $1/\tau_{\mathrm{ ee}}$  can only be reduced to the form quoted in Eq.~(\ref{tau_gen}). It should be stressed that the electron masses occurring in all equations of this section should be understood as {\em bare} rather than renormalized masses. This follows from the derivation of the BE in the Keldysh technique using the Migdal-Eliashberg approximation.

\end{widetext}


\begin{thebibliography}{100}



\bibitem{landau:36} L. D. Landau and I. J. Pomeranchuk, Phys. Z. Sowjetunion {\bf 10}, 649 (1936); Zh. Eksp. Teor. Fiz. {\bf 7}, 379 (1937).

\bibitem{abrikosov} A. A. Abrikosov, {\em Fundamentals of the Theory of Metals}, (North-Holland, Amsterdam, 1988). 

\bibitem{pomeranchuk:58} I. J. Pomeranchuk, Sov. Phys. JETP {\bf 8}, 361 (1958).

\bibitem{bass:90} J. Bass, W. P. Pratt, P. Schroeder, Rev. Mod. Phys. {\bf 62}, 646 (1990).

\bibitem{debye54}
 P. P. Debye and E. M. Conwell, Phys. Rev. {\bf 93}, 693 (1954).
 
 \bibitem{baber}  W. G. Baber, 
Proc. Roy. Soc. London {\bf 158}, 383 (1937).

\bibitem{levinson}V. F. Gantmakher and Y. B. Levinson, {\em Carrier
Scattering in Metals and Semiconductors} (North-Holland, Amsterdam,
1987). 

\bibitem{appel78}J. Appel and A. W Overhauser, \prb\;  {\bf 18}, 758 (1978).

\bibitem{gurzhi_eph}
R. N. Gurzhi, A. I. Kopeliovich, and S. B. Rutkevich, JETP Lett. {\bf 32}, 336 (1980).

\bibitem{gurzhi_ee} R. N. Gurzhi, A. I. Kopeliovich, and S. B. Rutkevich, JETP Lett. {\bf 56}, 159 (1982); c) Adv. Phys. {\bf 36}, 221 (1987).

\bibitem{maebashi97_98} a) H. Maebashi and H. Fukuyama, J. Phys.
Soc. Japan \textbf{66}, 3577 (1997); b) \textit{ibid.} \textbf{67},
242 (1998).

 \bibitem{rosch05} a) A. Rosch and P. C. Howell, \prb\; {\bf 72}, 104510 (2005); b) A. Rosch, Ann. Phys. {\bf 15}, 526 (2006).

\bibitem{maslov1} D. L. Maslov, V. I. Yudson, and A. V. Chubukov
Phys. Rev. Lett. {\bf 106}, 106403 (2011).

\bibitem{pal} H. K. Pal, V. I. Yudson, and D. L. Maslov, 
Phys. Rev. B {\bf 85}, 085439 (2012).
\bibitem{fradkin10}  E. Fradkin, S. A. Kivelson, M. J. Lawler, J. P. Eisenstein, A. P. Mackenzie, Ann. Rev. Cond. Matt. Phys. {\bf 1}, 153 (2010).

\bibitem{hmm}  J. A. Hertz,  Phys. Rev. B {\bf 14}, 1165 (1976); T. Moriya, {\em Spin Fluctuations in Itinerant Electron Magnetism} (Spinger-Verlag, Berlin, New York, 1985);  A. J. Millis, Phys. Rev. B {\bf 48}, 7183 (1993).

\bibitem{dzero04} M. Dzero and L. P. Gor'kov, \prb\; {\bf 69}, 092501 (2004); D. L. Maslov and A. V. Chubukov, {\em ibid.} {\bf 79}, 075112 (2009).

\bibitem{pepin06} a) A. V. Chubukov, C. P{\'e}pin, and J. Rech, Phys. Rev. Lett. {\bf 92}, 147003 (2004); b) J. Rech, C. P{\'e}pin, and A. V. Chubukov,  \prb\; \textbf{74}, 195126 (2006).

\bibitem{schofield:99}  A. J. Schofield,  Contemp. Phys. {\bf 40}, 95 (1999).

\bibitem{anna:07} L. Dell'Anna and W. Metzner, Phys. Rev. Lett. {\bf 98}, 136402 (2007) 
[Phys. Rev. Lett. {\bf 103}, 159904 (2009) (E)].

\bibitem{exp_fm} F. M. Grosche, C. Pfleiderer, G. J. McMullan, G. G. Lonzarich, 
and N. R. Bernhoeft, 
Physica B {\bf 206+207},  20 (1995); M. Nicklas et al.,
M. Brando, G. Knebel, F. Mayr, W. Trinkl, and A. Loidl,
Phys. Rev. Lett. {\bf 82}, 4268 (1999); P. G. Niklowitz,
F. Beckers, G. G. Lonzarich, G. Knebel, B. Salce, J. Thomasson, N. Bernhoeft, D. Braithwaite, and J. Flouquet,
Phys. Rev. B {\bf 72}, 024424 (2005).

\bibitem{physkin} E. M. Lifshitz and L. P. Pitaevskii, {\em Physical Kinetics}, (Oxford, Pergamon Press, 1981).

\bibitem{rammer}  J. Rammer and H. Smith, Rev. Mod. Phys. {\bf 58}, 323 (1986).

\bibitem{prange:64} R. E. Prange and L. P. Kadanoff, Phys. Rev. {\bf 134}, A566 (1964).

\bibitem{graf:93} See, e.g., M. J. Graf, D. Rainer, and J. A. Sauls, Phys. Rev. B {\bf 47}, 12089 (1993).

\bibitem{eliashberg_breakdown} S.-S. Lee, Phys. Rev. B {\bf 80}, 165102 (2009); M. A. Metlitski and S. Sachdev, Phys. Rev. B {\bf 82}, 075127 (2010); D. F. Mross, J. McGreevy, H. Liu, and T. Senthil, Phys. Rev. B {\bf 82}, 045121 (2010);  A.V. Chubukov,  Physics {\bf 3}, 70 (2010). 

\bibitem{sturman84} B. I. Sturman, 
Sov. Phys.-Uspekhi \textbf{27} 881 (1984) {[}Usp. Fiz. Nauk \textbf{144},
497 (1984){]}. 

\bibitem{comment_ei} Since Ref.~\onlinecite{sturman84} is not
widely available, we take a liberty to reproduce its main message
here: Some sources quote a different
form of $I_{\mathrm{ei}}$: $I_{\mathrm{ei}}^{*}=\sum_{\bkp}w_{\bkp,\bk}f_{\bkp}(1-f_{\bk})-w_{\bk,\bkp}f_{\bk}(1-f_{\bkp})$.
This form is incorrect: it does not follow from a microscopic theory
and falsely predicts a possibility of non-linear evolution of the
distribution function in a linear system. Under a more restrictive
assumption of microreversibility, the non-linear terms
cancel out and $I_{ei}^{*}$ becomes correct. 

\bibitem{comment_RTA} Strictly speaking, the second term in the collision integral
should contain the distribution function averaged over the directions of $\bk$ rather than the
equilibrium distribution $n_{\bk}$. However, this difference is immaterial to linear order in ${\bf E}$.

\bibitem{blokh} M. D. Blokh and L. I. Magarill, Sov. Phys. Solid
State \textbf{22}, 1327 (1080) {[}Fiz. Tverd. Tela \textbf{22}, 2279
(1980){]}. 

\bibitem{comment_compensated}
A well-known case when the resistivity is finite only due to normal {\em ee} collisions is a perfectly compensated semi-metal.~ \cite{levinson}
Undoped graphene is a special case of a compensated system with zero Fermi energy,  where the conductivity is finite (and universal) at $T=0$ even in the absence of any scattering.~\cite{mishchenko07}

\bibitem{maslov04} S.  Gangadharaiah, D. L. Maslov, A. V. Chubukov, and L. I. Glazman, Phys. Rev. Lett. \textbf{94},
156407 (2005); A. V. Chubukov, D. L. Maslov, S. Gangadharaiah,
and L. I. Glazman, Phys. Rev. B \textbf{71}, 205112 (2005).

\bibitem{lyakhov03} A. O. Lyakhov and E. G. Mishchenko, Phys. Rev.
B \textbf{67}, 041304 (2003).

\bibitem{gurzhi95} R. N. Gurzhi, A. N. Kalinenko, and A. I. Kopeliovich, Phys. Rev. B \textbf{52}, 4744 (1995).

\bibitem{fu} L. Fu, Phys. Rev. Lett. \textbf{103}, 266801 (2009).

\bibitem{comment3} More than two points of intersection may be possible for large momentum transfers of $q\approx 2k_F$. However, this leads to higher order terms in $\Delta$.

\bibitem{herring56} C. Herring and E. Vogt, Phys. Rev. {\bf 93}, 693 (1956)

\bibitem{keyes58} R. Keyes, J. Phys. Chem. Solids {\bf 6}, 1 (1958).

\bibitem{gantmakher78} V. F. Gantmakher and Y. B. Levinson, JETP {\bf 47}, 133 (1978).

\bibitem{ziman} T. J. Ziman, {\em Electrons and Phonons: The Theory
of Transport Phenomena in Solids}, (Oxford, 2001). 

\bibitem{altshuler:85}  B. L. Altshuler and A. G. Aronov, in {\em Electron-Electron Interactions in
Disordered Systems}, edited by A. L. Efros and M. Pollak (Elsevier,
1985), p. 1.

\bibitem{zala:01}  G.\ Zala, B.\ N.\ Narozhny, and I.\ L.\ Aleiner, Phys. Rev. B {\bf 64}, 214204 (2001).

\bibitem{paul:05}  I. Paul, C. P{\'e}pin, B. N. Narozhny , and D. L. Maslov, Phys. Rev. Lett. {\bf 95}, 017206 (2005); I. Paul, \prb\; {\bf 77}, 224418 (2008).

\bibitem{kravchenko} B. Spivak, S. V. Kravchenko, S. A. Kivelson, and X. P. A. Gao, Rev. Mod. Phys. {\bf 82}, 1743 (2010).

\bibitem{gurzhi64} R.N. Gurzhi, Sov. Phys. JETP {\bf 17}, 521 (1964).

\bibitem{hruska02} M. Hruska and B. Spivak, \prb \textbf{65}, 033315
(2002).

\bibitem{fluid_dynamics} L. D. Landau and E. M. Lifshits, \textit{Fluid
Mechanics} (Pergamon Press, Oxford 1987).

\bibitem{phys_kin} Ref.~\onlinecite{physkin}, p. 62.  

\bibitem{metamagn}  R. S. Perry, K. Kitagawa, S. A. Grigera, R. A. Borzi,
A. P. Mackenzie, K. Ishida, and Y. Maeno, 
Phys. Rev. Lett. {\bf 86}, 2661 (2001);
S. A. Grigera, R. S. Perry, A. J. Schofield, M. Chiao,
S. R. Julian, G. G. Lonzarich, S. I. Ikeda, Y. Maeno, A. J. Millis, A. P.
Mackenzie, 
Science {\bf 294},
329 (2001); S. A. Grigera,P. Gegenwart, R. A. Borzi, F. Weickert,
A. J. Schofield, R. S. Perry, T. Tayama, T. Sakakibara,
Y. Maeno, A. G. Green, A. P. Mackenzie, Science {\bf 306}, 1154 (2004).

\bibitem{murzin98} S. S. Murzin, S. I. Dorozhkin, G. Landwehr, and
A. C. Gossard, JETP Lett. \textbf{67}, 113 (1998).

\bibitem{mishchenko07} A. W. W. Ludwig, M. P. A. Fisher, R. Shankar, and G. Grinstein, Phys. Rev. B {\bf 50}, 7526 (1994);
E. G. Mishchenko, \prl\; \textbf{98}, 216801
(2007).




\end{thebibliography}
\end{document}